\documentclass[5p]{elsarticle}
\usepackage[utf8]{inputenc}
\usepackage[english]{babel}
\usepackage[T1]{fontenc} 
\usepackage{amsmath,amssymb}
\usepackage{upgreek}
\usepackage{graphicx}
\usepackage{bm}

\usepackage{pdfpages}

\usepackage[separate-uncertainty=true,multi-part-units=single, allow-number-unit-breaks]{siunitx}\usepackage{bm}
\usepackage[version=4]{mhchem} 
\usepackage{soul, color}

\usepackage{hyperref}
\usepackage{natbib}
\usepackage{tabulary, booktabs} 

\begin{document}

\begin{frontmatter}

\author[add1]{Jeppe Christiansen \fnref{1}}
\author[add2]{Joakim Vester-Petersen \fnref{1}}
\author[add1,add3]{Søren Roesgaard}
\author[add1]{Søren H. Møller}
\author[add4]{Rasmus E. Christiansen}
\author[add4]{Ole Sigmund}
\author[add2]{S{\o}ren P. Madsen}
\author[add1,add3]{Peter Balling}
\author[add1,add3]{Brian Julsgaard}
\ead{brianj@phys.au.dk}

\fntext[1]{Contributted equally to this work}

\address[add1]{Department of Physics and Astronomy, Aarhus University, Ny Munkegade 120, DK-8000 Aarhus C, Denmark}

\address[add2]{Department of Engineering, Aarhus University, Inge Lehmanns Gade 10, DK-8000 Aarhus C, Denmark}

\address[add3]{Interdisciplinary Nanoscience Center (iNANO), Aarhus University, Gustav Wieds Vej 14, DK-8000 Aarhus C, Denmark}

\address[add4]{Department of Mechanical Engineering, DTU, Nils Koppels Allé 404, DK-2800 Kgs. Lyngby, Denmark}

\title{Strongly enhanced upconversion in trivalent erbium ions by tailored gold nanostructures: toward high-efficient silicon-based photovoltaics}

\begin{abstract}
Upconversion of sub-band-gap photons constitutes a promising way for improving the efficiency of silicon-based solar cells beyond the Shockley-Queisser limit. \SIrange{1500}{980}{\nano\meter} upconversion by trivalent erbium ions is well-suited for this purpose, but the small absorption cross section hinders real-world applications. We employ tailored gold nanostructures to vastly improve the upconversion efficiency in erbium-doped \ce{TiO2} thin films. The nanostructures are found using topology optimization and parameter optimization and fabricated by electron beam lithography. In qualitative agreement with a theoretical model, the samples show substantial electric-field enhancements inside the upconverting films for excitation at \SI{1500}{\nano\meter} for both s- and p-polarization under a wide range of incidence angles and excitation intensities. An unprecedented upconversion enhancement of \num{913(51)} is observed at an excitation intensity of \SI{1.7}{\watt\per\square\centi\meter}. We derive a semi-empirical expression for the photonically enhanced upconversion efficiency, valid for all excitation intensities. This allows us to determine the upconversion properties needed to achieve significant improvements in real-world solar-cell devices through photonic-enhanced upconversion.
\end{abstract}

\begin{keyword}
Upconversion of sub-band gap photons, high-efficient photovoltaics, photonic enhancement, topology optimization
\end{keyword}

\end{frontmatter}

\section{Introduction}
Photon upconversion \cite{Bloembergen1959, Brown1969, Auzel1984, Ovsyakin1966}, the photoluminescence process in which the emission wavelength is shorter than the excitation wavelength, has exciting applications in many fields such as bio-imaging \cite{Li2006, Chatterjee2010}, anti-counterfeiting \cite{Liu2011, You2015}, and not least in improving the efficiency of solar cells \cite{Shalav2005, Goldschmidt2015, Balling2018}. Different mechanisms are known to be able to upconvert light, but among the most promising is upconversion from trivalent lanthanide ions embedded in a glass or crystalline host material. The multitude of 4f-4f electronic transitions in the lanthanide series allows for a wide range of upconversion wavelengths, and the real intermediate states provide the opportunity to upconvert incoherent and less intense light. However, the active 4f-4f transitions in lanthanide ions are dipole forbidden, causing low absorption and thus low external quantum efficiency \cite{Auzel2004}, which remains an obstacle, especially for applications in solar cells -- the main focus of this investigation.

Several pathways for enhancing the upconversion process have been proposed during the past decade, some of which focus on increasing the absorption by co-doping with a sensitizer \cite{Chen2015-3} while others utilize photonic enhancement through wave-guiding effects and Bragg stacks \cite{Johnson2011, Hofmann2018} or through surface-plasmon resonances \cite{Wu2014, Park2015}. Common for all photonic enhancements of lanthanide-based upconversion is that they affect the upconversion process in two ways: First, they can influence the relaxation rates both positively, through enhanced coupling to the radiation field, and negatively, through quenching via ohmic heating in the surrounding material \cite{Wu2014, Park2015}. Second, they can cause an increase in the electric field at the location of the lanthanide ions and hence also increase the absorption, which in turn will improve the upconversion efficiency \cite{Wu2014, Park2015}.

\begin{figure*}[t!]
\centering
\includegraphics[scale=1]{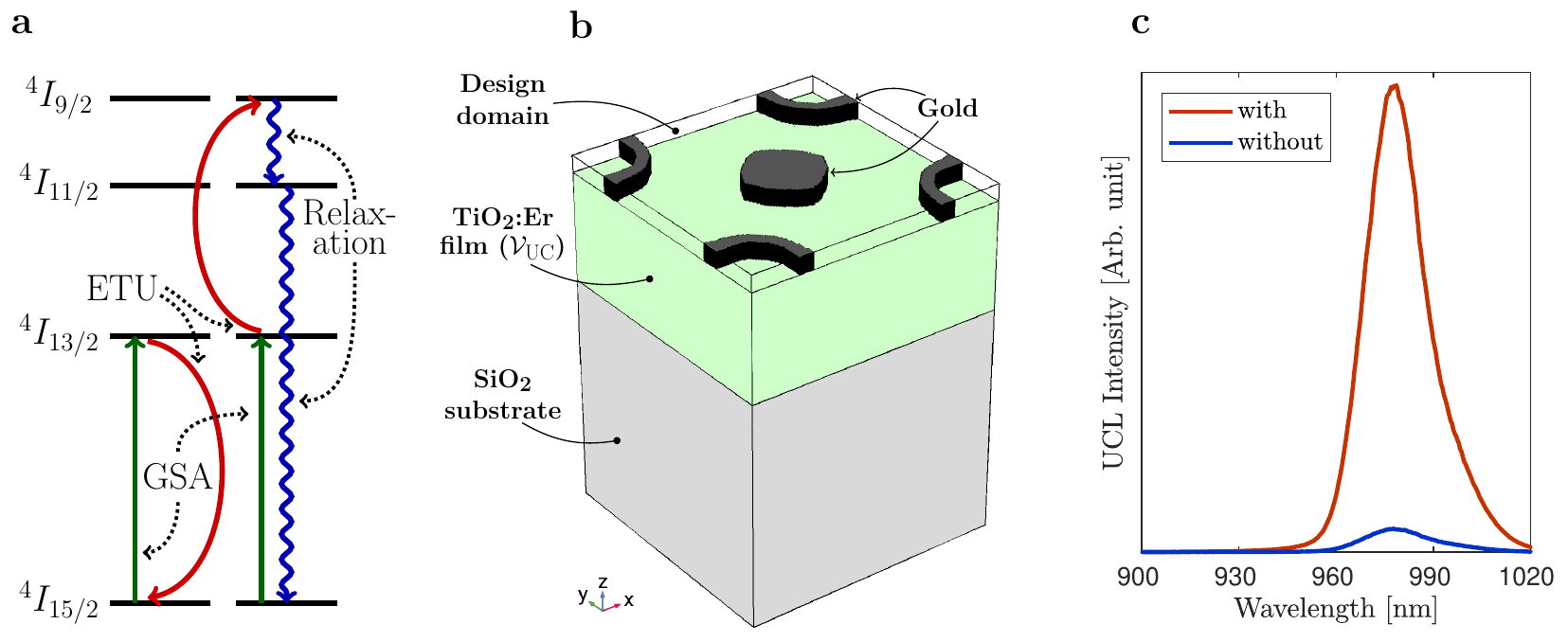} 
\caption{In a), the excitation process is sketched in two erbium ions. Initial ground-state absorption excites two erbium ions to the  intermediate state $^4I_{13/2}$ indicated with green vertical arrows. In $^4I_{13/2}$, the two ions can interact with each other via an energy-transfer-upconversion process illustrated with red curved arrows, which leaves one ion in the ground state and the other in the doubly-excited state: $^4I_{9/2}$. From here, a two-step relaxation process indicated with curly blue arrows is most likely to occur. First, the ion will predominantly relax to the $^4I_{11/2}$ by a nonradiative relaxation due to the small energy difference between the $^4I_{9/2}$ and $^4I_{11/2}$ levels. Second, the ion will relax to the ground state either nonradiative (loss channel) or radiative by emitting a photon of higher energy than was initially absorbed. In b), the model setup is shown. The gold nanostructure (indicated in black) is placed in a \SI{50}{\nano\meter}-high design domain on top of a \SI{320}{\nano\meter}-thick film of erbium-doped \ce{TiO2} (green) on top of a \SI{0.5}{\milli\meter} \ce{SiO2} substrate (gray). In c), the upconversion-luminescence is exemplified by two spectra showing the measured upconversion luminescence with (red) and without (blue) the gold nanostructure present on the film surface.}
\label{fig:overview}
\end{figure*}  

This work considers an upconverting material consisting of erbium ions (\ce{Er^3+}) doped into a \ce{TiO_2} thin film (\ce{TiO_2}:\ce{Er}). The \ce{Er^3+} ions are able to absorb light at wavelengths in the vicinity of \SI{1500}{\nano\meter} \cite{Goldschmidt2015} and in turn emit upconverted electromagnetic (EM) radiation at \SI{980}{\nano\meter} due to the process sketched in Fig.~\ref{fig:overview}a. Carefully designed periodic gold nanostructures are placed on the surface of the \ce{TiO_2}:\ce{Er} film, as sketched in Fig.~\ref{fig:overview}b, in order to couple incident \SI{1500}{\nano\meter} EM radiation into the film and thereby concentrate it. These nanostructures lead to an increased emission of upconverted radiation, which is directly measurable, as exemplified in Fig.~\ref{fig:overview}c. Since the wavelength of the upconverted light falls within the absorption range of crystalline silicon solar cells, the upconversion process has the potential to improve the efficiency of such solar cells, and the ability to enhance the upconversion process constitutes an important step toward this goal.

There are two main results of this work: First, it is demonstrated that a well-chosen film geometry, which supports wave-guided modes of \SI{1500}{\nano\meter} EM radiation, in combination with light concentration, facilitated by the well-chosen gold nanostructures, leads to an unprecedented enhancement of the upconversion-luminescence (UCL) yield. Second, it is demonstrated how the numerical calculation of electric fields within the upconverting film in combination with a recently developed analytical model for the upconversion process can account rather accurately for the observed UCL enhancement for varying intensities, incidence angles, and polarization orientations. As a result, the relation between the measured enhancements and the concentration factor of the incident radiation becomes apparent, which enables a justified prediction of the steps required before such upconverting materials can reach efficiencies relevant for improving the performance of solar cells.

\section{Theory}  
The upconversion process has a nonlinear dependence on the excitation intensity, which saturates as the intensity is increased. For two-photon upconversion, the upconversion-intensity dependence will saturate from a quadratic dependence in the low-excitation limit toward a linear dependence in the high-excitation limit \cite{Pollnau2000}. In a recent paper \cite{Christiansen2019-1}, we have shown that for a bare upconverting film the intensity dependence of the UCL yield, $Y_{\mathrm{UCL}}$, is given by
\begin{equation}
  \label{eq:Ysat}
  \begin{split}
    Y_{\mathrm{UCL}} &= A\left\lbrace 1 +
       \ln
       \left[\frac{\sqrt{1 + 2 I/I_{\mathrm{sat}}} + 1}{2}\right] \right. \\
     &\left.
        + \frac{I}{2I_{\mathrm{sat}}} - \sqrt{1+ 2I/I_{\mathrm{sat}}} \right\rbrace \\
        &\equiv A f  \left( I/I_{\mathrm{sat}} \right)
  \end{split}
\end{equation}
where $A$ is an amplitude factor, accounting for material parameters and detection efficiency, and $I_{\mathrm{sat}}$ is the excitation intensity where the upconversion process starts to saturate. A good upconverter has a low saturation intensity, exemplified by the fact that an increased absorption cross section decreases the saturation intensity, whereas an increased nonradiative relaxation will increase the saturation intensity \cite{Christiansen2019-1}. When gold nanostructures are added to the film surface, causing the enhanced UCL, as exemplified in Fig.~\ref{fig:overview}c, the model becomes more involved. However, the UCL enhancement is straightforward to determine experimentally by measuring the total UCL yield (area under each curve in Fig.~\ref{fig:overview}c) with and without the nanostructures present on the surface of the upconverter (red and blue curves, respectively) at the same excitation conditions. In other words, the enhancement is a very convenient tool for characterizing the impact of photonic nanostructures, and it is, therefore, worthwhile to seek a theoretical understanding of this enhancement. By combining the above-mentioned analytical model with simulated electric-field distributions inside the film, the theoretical UCL enhancement, $L_\mathrm{UCL}$, defined as the ratio of $Y_\mathrm{UCL}$ with and without gold nanostructures present on the film surface, can be calculated as
\begin{equation}
\label{eq:Lsat}
L_{\mathrm{UCL}} = \frac{
{\displaystyle \int_{\mathrm{\mathcal{V}_{\mathrm{UC}}}} } f\left(
\frac{|\mathbf{\tilde{E}}|^2}
{\overline{|\mathbf{\tilde{E}}_{\mathrm{s}}|^2}} \frac{I}{I_\mathrm{sat}} \right) \mathrm{d}V
}
{
{\displaystyle \int_{\mathrm{\mathcal{V}_{\mathrm{UC}}}} } f\left(
\frac{|\mathbf{\tilde{E}}_\mathrm{b}|^2}
{\overline{|\mathbf{\tilde{E}}_{\mathrm{s}}|^2}} \frac{I}{I_\mathrm{sat}} \right) \mathrm{d}V
},
\end{equation}   
where $f$ is the saturation function defined in Eq.~\eqref{eq:Ysat}, and the integration is taken over the volume, $\mathcal{V}_{\mathrm{UC}}$, of the upconverting film below one unit cell of the gold nanostructures, see Fig.~\ref{fig:overview}b. All electric fields in this expression are simulated according to the relevant experimental conditions (i.e., angle of incidence and polarization), and $\mathbf{\tilde{E}}$ and $\mathbf{\tilde{E}}_{\mathrm{b}}$ correspond to the fields in the presence and absence of gold nanostructures, respectively. Due to the linearity of the absorption process, the squared electric field inside the upconverting film is proportional to the intensity, $I$, of the incoming laser beam and hence only needs to be simulated once. The correct scaling is calibrated by reference to the laser-beam intensity, $I_{\mathrm{sat}}$, which drives the upconverter material into saturation, and the corresponding simulated electric field, $\mathbf{\tilde{E}_{\mathrm{s}}}$, for the experimental conditions of this calibration measurement. The bar over $|\mathbf{\tilde{E}_{\mathrm{s}}}|^2$ in Eq.~\eqref{eq:Lsat} denotes volume-averaging over $\mathcal{V}_{\mathrm{UC}}$. Variations in the relaxations rates, e.g., quenching, are neglected since these effects will only be significant for the small population of erbium ions in close vicinity of the gold nanostructure, in agreement with Ref.~\cite{Eriksen2019}. A detailed derivation of Eq.~\eqref{eq:Lsat} is available in Sec.~4.1 of the supporting information; later in this manuscript, we shall explain how the enhancement relates to the more general description of light concentration and improvement of photovoltaics.

Fig.~\ref{fig:overview}b shows the model setup for the simulation of the electric fields consisting a (x,y)-periodic unit cell of a \SI{320}{\nano\meter}-thick \ce{TiO_2}:\ce{Er} film and \SI{50}{\nano\meter}-tall gold nanostructures placed on top, in the design domain. The goal is to obtain nanostructure designs, which enable efficient coupling of the incident light to guided modes in the \ce{TiO_2}:\ce{Er} film, and thereby, enhance the light intensity inside the film. The first step in designing such structures is to calculate the electric field accurately. Assuming nonmagnetic, linear, and isotropic materials, the Maxwell equations can be recast into the time-harmonic vector-wave equation \cite{Novotny2006}
\begin{equation}
\label{eq:wave_equation}
\nabla \times (\nabla \times \bm{E} ) - \omega^2\mu_0\epsilon(\bm{r}) \bm{E} = 0,
\end{equation} 
where $\omega$ is the angular frequency, $\mu_{0}$ is the free space permeability, and $\epsilon(\bm{r})$ is the position-dependent (complex) electric permittivity, with $\bm{r}=(x,y,z)$ denoting the spatial position. Eq.~\eqref{eq:wave_equation} is solved numerically using the finite element method \cite{Jin2014}. Once the electric field can be calculated, the next step is to optimize the nanostructure design to enhance the UCL yield, which is numerically evaluated through an appropriate merit function (to be introduced). The approach taken here, known as topology optimization, is to define a material distribution in the design domain and maximize the merit function by optimizing this distribution. This is achieved by changing $\epsilon(\bm{r})$ in Eq.~\eqref{eq:wave_equation}, through a spatially dependent nonphysical field $\rho \in [0,1]$ and expressing the permittivity as $\epsilon(\eta(\rho),\kappa(\rho))$ \cite{RChristiansen2019}, where $\eta$ and $\kappa$ are the refractive index and extinction coefficient, respectively. The design domain is divided into equally sized voxels (cubes), and a single $\rho$-value is assigned to each voxel with a value of $\rho = 0$ corresponding to air and $\rho = 1$ to gold. Since $\rho$ is allowed to be continuous, gradient-based optimization algorithms can be used to solve the optimization problem efficiently. Any nonphysical material mixes (where $\rho\neq 0$ and $\rho\neq 1$) are gradually removed over the course of the optimization using standard penalization tools \cite{Guest2004,Wang2011,Christiansen2015}, eventually forcing $\rho$ toward either 0 or 1. This introduction outlines the general idea behind topology optimization. For an in-depth presentation of topology optimization applied to nano and microscale electromagnetism, the reader is referred to Ref.~\cite{Jensen2010}.

We choose to optimize the upconversion performance in the middle of the excitation regime, between the linear and quadratic intensity dependence, that is, for a cubic dependence on the electric field. The optimization problem is therefore formulated with the merit function \cite{Madsen2015,Johannsen2015,VesterPetersen2017,VesterPetersen2018}
\begin{equation}
\Phi_{i,j}  = \frac{\int_{\mathcal{V}_{\mathrm{UC}}}\vert\bm{\tilde{E}}^i(\bm{\rho},\lambda_j)\vert^3~\mathrm{d}V} 
{\int_{\mathcal{V}_{\mathrm{UC}}}\vert\bm{\tilde{E}}^i_{\mathrm{b}}(\bm{0},\lambda_j)\vert^3~\mathrm{d}V},
\label{eq:objective_function}
\end{equation}
where $i$ and $\lambda_j$ parameterize the polarization and wavelength of the incoming field, respectively. In order to promote a design with a reasonable robustness against slight experimental fabrication errors, the optimization problem is formulated as a \textit{min-max} problem such that $i \in {\mathrm{s},\mathrm{p}}$ and $\lambda_j \in \{\SI{1490}{\nano\meter},\SI{1500}{\nano\meter},\SI{1510}{\nano\meter}\}$ are considered simultaneously. By minimizing the worst performance in the set of all realizations, $\left\{-\Phi_{i,j}\right\}$, the sensitivity toward variation in polarization and wavelength, or equivalently size variations in the realized nanostructure array, is decreased. 

\section{Materials and methods}
\subsection{Numerical calculations}
The topology optimization was performed with a voxel size of ($5 \times 5 \times 5$) \si{\cubed\nano\meter} and with a 2D $\rho$-field, extruded to a height of \SI{50}{\nano\meter} to ensure that the designs can be fabricated using electron-beam lithography (EBL). A unit-cell period of \SI{780}{\nano\meter} was chosen to allow for efficient coupling to waveguide modes in the film \cite{VesterPetersen2018}. Since the optimization problem is nonconvex, different geometries may emerge for different initial $\rho$'s. A set of calculations have been made with a random initial material distribution, resulting in a (smaller) set of converged geometries. These geometries have been numerically characterized, and we have chosen one, which shows robustness toward a change in incidence angle, while still yielding high enhancements. In this procedure, we have on purpose excluded designs such as a simple disk grating, which may work better for a particular wavelength and angle of incidence, but the extreme narrow-band response makes such structures irrelevant for developing devices that must utilize the broad-bandwidth solar radiation. A movie showing the course of the topology-optimization process leading to the chosen design is found in the supplementary material, and the chosen design, denoted P780*, can be seen in Fig.~\ref{fig:designs}a.

Inspired by the simple square-ring structure of P780*, we have shape-optimized designs consisting of a square and a ring by tuning the square width, ring radius, and ring thickness using a derivative-free optimization algorithm (Nelder-Mead \cite{Arora2017}) to maximize the merit function in Eq.~\eqref{fig:RecordEnhancement}, as for P780*. Unit-cell periods of \SI{780}{\nano\meter}, \SI{800}{\nano\meter}, and \SI{1000}{\nano\meter} were chosen and all other parameters kept the same as for the topology-optimized sample P780*. The resulting three geometries denoted P780, P800, and P1000 can be seen in Fig.~\ref{fig:designs}b-d.

More details of the numerical calculations are available in Sec.~1 of the supporting information.

\subsection{Sample fabrication}
\label{subsec:sample_fabrication}
Starting with a \SI{0.5}{\milli\meter}-\ce{SiO_2} substrate, the \ce{TiO_2}:\ce{Er} was deposited using a radio-frequency (RF) magnetron-sputtering system. The targets were commercially produced from powders of \ce{TiO_2} and \ce{Er_2O_3} at an erbium concentration of 5.1 at.\%. The sputtering process was conducted in an argon atmosphere with \SI{2}{\percent} oxygen at a pressure of \SI{3}{\mmHg}. The sputtering was done with a fixed RF power of \SI{100}{\watt} and with the substrate temperature fixed at \SI{350}{\degreeCelsius}. These conditions were found to minimize unwanted nonradiative relaxations in the thin films \cite{Lakhotiya2018-JAP}. The deposition time was calibrated to achieve a film thickness of \SI{320}{\nano\meter}.

The gold nanostructures were fabricated by EBL in combination with physical-vapor deposition on an ($1.2 \times 1.2$) \si{\milli\meter\square} area of the thin films to allow upconversion measurements on and off the gold nanostructures on the same sample by moving the laser spot on the surface, and to minimize the fabrication time. The EBL process was corrected for proximity effects using the procedure described in Ref.~\cite{Eriksen2018} to obtain a better resolution. The fabricated gold nanostructure had sidewalls with an inclination angle of \ang{75} as opposed to vertical sides (\ang{90}) assumed in the topology optimization. For more information on the EBL, the reader is referred to Sec.~2.2 of the supporting information.

\subsection{Optical diffraction measurements}
\label{subsec:Opt_Difrac}
After EBL fabrication, the mean unit-cell period of the nanostructures was measured using optical diffraction. The measurements were conducted by transmitting a helium-neon laser onto the nanostructure facing a screen where the created diffraction pattern could be seen. The unit-cell period was then calculated from Bragg's law using: the wavelength of the laser, the distance to the wall, and the distance from the zeroth to the first-order diffraction peak on the screen.

This investigation showed an asymmetry in the horizontal and vertical directions (defined by the SEM image), which leads to ambiguities when measuring the optical properties at different polarization orientations and angles of incidence. We have chosen to present the results with the electric field of the excitation source, parallel to the vertical period of the nanostructures here, while the case with the electric field parallel to the horizontal period can be seen in Fig.~S2 in the supporting information.

A thorough explanation of these measurements and results are available in Sec.~3.1 of the supporting information.

\subsection{Upconversion luminescence measurements}
\label{subsec:UCLMeas}
The UCL yield was measured by illuminating the samples with \SI{1500}{\nano\meter} laser light and recording the luminescence with an integrated spectrometer and CCD camera. The UCL-enhancement measurements were conducted with the samples placed in an integrating sphere with a diameter of \SI{150}{\milli\meter} to obtain identical collection efficiencies for all used geometries. The UCL yield is determined by integrating across the \SI{980}{\nano\meter} peak shown in Fig.~\ref{fig:overview}c. The enhancement is defined as the ratio of the UCL yield on and off the gold nanostructures.

The UCL enhancements were measured at varying angles of incidence between \ang{0} and \ang{25}, for both s and p-polarization, and for two excitation intensities, \SI{323 \pm 97}{\watt\per\square\centi\meter} and \SI{5.8 \pm 0.5}{\watt\per\square\centi\meter}, obtained using two different laser-beam areas. The rather high uncertainty in the high-excitation intensity originates from the laser-beam area, see Sec.~3.3 of the supporting information. The enhancements with the horizontal period parallel to the electric field are shown in Fig.~S2 of the supporting information.

To allow comparison with Eq.~\eqref{eq:Lsat}, the saturation intensity of the sample was measured at a \ang{50} angle of incidence and a p-polarized excitation, and by fitting UCL intensity-dependence data to Eq.~\eqref{eq:Ysat}, we have determined the corresponding saturation intensity to $I_{\mathrm{sat}} = \SI{20 \pm 6}{\watt\per\square\centi\meter}$, see data and fit in Fig.~S3 of the supporting information.

A thorough account of the UCL measurements, including the determination of the saturation intensity, is available in Sec.~3.2 of the supporting information.

\section{Results}
\label{sec:results}
\begin{figure*}[!ht]
\centering
\includegraphics[width = \textwidth]{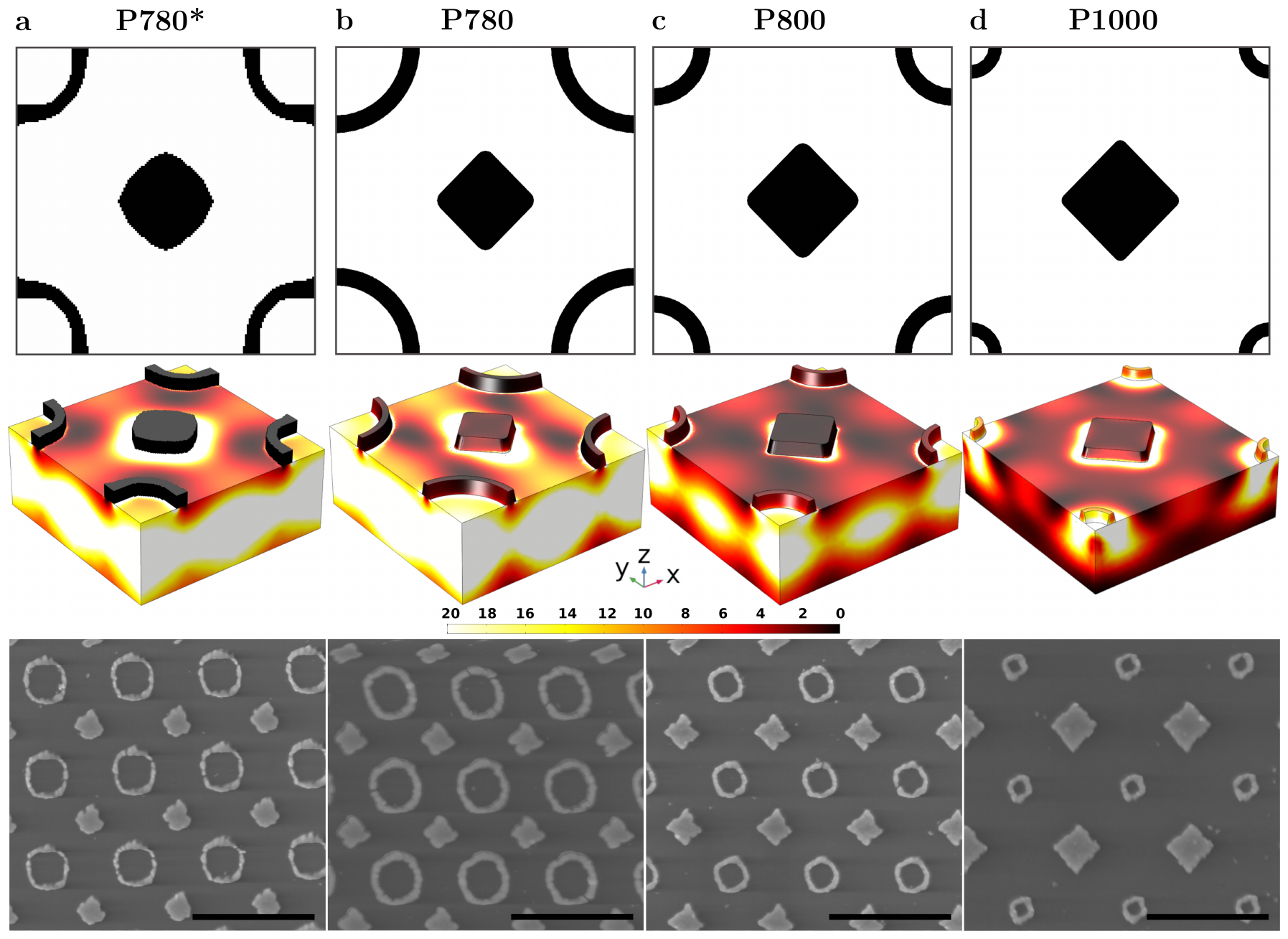}
\caption{Optimized and fabricated designs a) P780*, b) P780, c) P800, and d) P1000. In the top row, the unit-cell designs are shown with white representing air and black representing gold, with the gray frame indicating the unit-cell boundary.  In the middle row, the polarization-averaged energy-density enhancement in the film, $1/2 \sum_{\mathrm{s},\mathrm{p}} |\bm{E}|^2/|\bm{E}_{\mathrm{b}}|^2$, at $\lambda = \SI{1500}{nm}$ is shown. In the bottom row, SEM pictures of the EBL fabricated gold nanostructures are shown with the scale-bars indicating \SI{1}{\micro\meter}.}
\label{fig:designs}
\end{figure*}
In Fig.~\ref{fig:designs}a, the topology-optimized design is shown in the upper row with the electric-field enhancement and distribution underneath, and a scanning-electron microscope (SEM) image of the EBL produced sample in the lowest row. The samples presented in Fig.~\ref{fig:designs}b-d will be discussed later.
\begin{figure*}[ht!]
\centering
\includegraphics[scale=1]{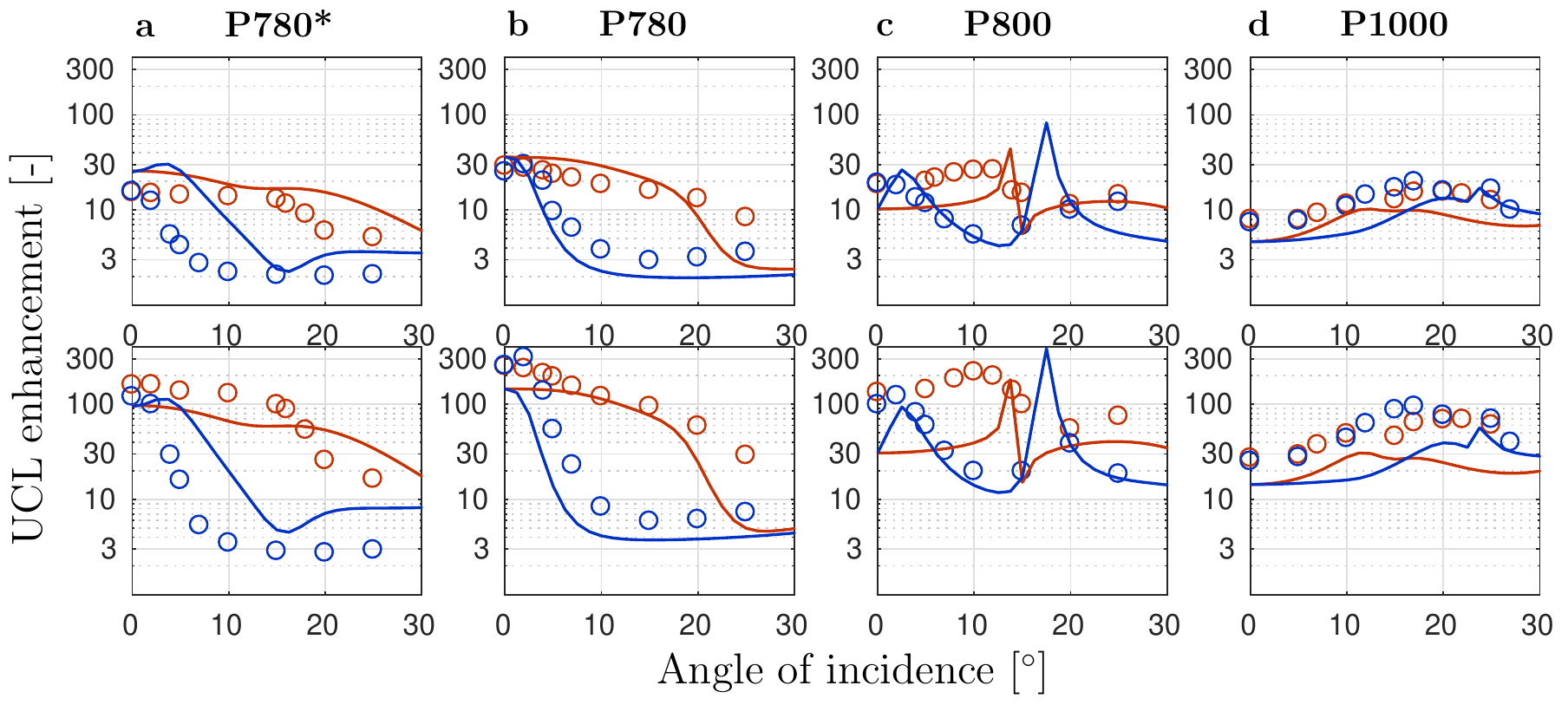}
\caption{UCL enhancements for all four nanostructure designs at high excitation intensity \SI{323}{\watt\per\square\centi\meter} (upper row), and low intensity \SI{5.8}{\watt\per\square\centi\meter} (lower row). The measurements are plotted by circles, and the calculated values are represented with solid curves. Red is used for the p-polarized case and blue for the s-polarized case.}
\label{fig:Enh_Y}
\end{figure*}
The measured UCL enhancements with the vertical period parallel to the electric field are presented in Fig.~\ref{fig:Enh_Y} with the high and low-intensity cases in the upper and lower panels, respectively. The theoretically predicted upconversion enhancements, $L_{\mathrm{UCL}}$, shown as solid curves in Fig.~\ref{fig:Enh_Y}, were calculated using Eq.~\eqref{eq:Lsat}. A reasonable qualitative agreement between the measurements and simulations is observed: the order of magnitude, as well as the trend of the variations with incidence angle and polarization, agree mostly. We stress that the calculated UCL enhancements are computed from only the simulated electric-field distribution, and the experimentally determined saturation intensity. Moreover, we emphasize that the electric-field calculations are made from the ideal unit-cell design with the periods and slanting angle scaled to the measured values, as explained in Sec.~\ref{subsec:sample_fabrication} and \ref{subsec:Opt_Difrac}. The fabrication imperfections observed by comparing the design and SEM images in the first and third row of Fig.~\ref{fig:designs}, respectively, are thus not accounted for in the calculations, and the model is hence only expected to be able to describe the general trend of the UCL enhancement caused by the nanostructures.

A significant UCL enhancement has been found for a wide range of incidence angles, especially for p-polarized excitation. This is in stark contrast to what is generally reported in the literature, where high UCL enhancements typically are coupled to a strong angular dependence \cite{Hofmann2018, Verhagen2009}. A broad angular acceptance is essential when considering the end goal of improving photovoltaics by concentrating solar radiation. Noteworthy is also the fact that the enhancements reported here are obtained in thick samples (\SI{320}{\nano\meter}) contrary to the case of previously reported high UCL enhancements, which were recorded from upconverting film thicknesses below \SI{25}{\nano\meter} \cite{Verhagen2009, Zhang2012}. Due to the low absorption cross section of erbium, thick upconverters are essential to achieve adequate external quantum efficiency. From the absorption cross section of erbium \cite{Miniscalco1991} and the general concentration level of erbium-doped upconverters the absorption length is on the order of \SI{1}{\milli\meter} at \SI{1500}{\nano\meter}, hence we will need even thicker films to achieve adequate external quantum efficiencies; however, the results reported here provide an initial step on the way to the realization of upconversion-enhanced solar cells.

The three additional designs in Fig.~\ref{fig:designs}b-d were found using parameter optimization to investigate how the unit-cell period affects the UCL yield, as they exhibit, based on simple grating-coupling condition, an efficient, a moderate, and a weak waveguide coupling at normal incidence, respectively. The additional designs also further allow us to validate the UCL model under different illumination conditions. The obtained designs have been fabricated, and their upconversion properties measured similarly to P780*. Looking at the field enhancements in the second row of Fig.~\ref{fig:designs}, it is clear that the field distribution of P780 (and P780*) resembles that of a waveguide mode with field enhancement spanning the entire film. In contrast, P1000 shows a field distribution with distinctly localized enhancements typically attributed to plasmonic resonances. In Fig.~\ref{fig:Enh_Y}, large UCL enhancements are observed both experimentally and numerically at normal incidence for P780 (and P780*), while the samples P800 and P1000 show significantly lower values at normal incidence and additionally have their peak efficiency at a nonzero angle of incidence. This observation can be attributed to the phase-matching criterion for coupling the incoming light to the waveguide mode \cite{VesterPetersen2018}. The criterion is not exactly met at normal incidence but is instead fulfilled better at some nonzero angle. The UCL enhancements in Fig.~\ref{fig:Enh_Y} generally show a reasonable agreement between measured and calculated values in all cases, except in P800 at angles where the numerical data show strong resonances, which may not be present in the fabricated samples or is simply not captured due to the finite number of measuring angles. This generally good agreement between measurements and calculations is further backed by measured and calculated extinction cross sections available in Fig.~S5 of the supporting information.

\begin{figure}[t!]
\centering
\includegraphics[scale = 1]{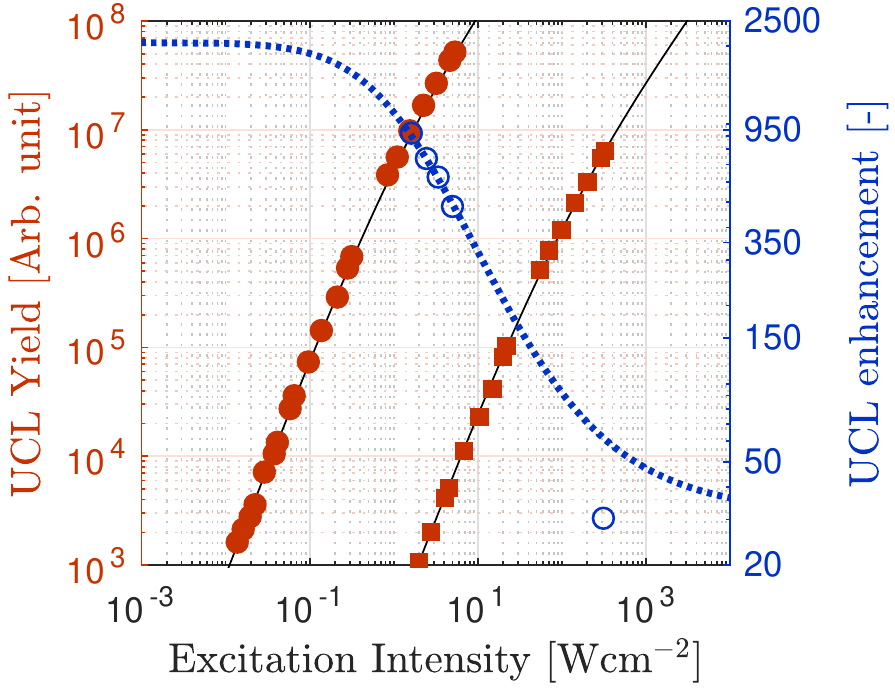}
\caption{The UCL enhancement for the P780 sample is plotted against excitation intensity in blue open-faced points. The red circles and squares are intensity-dependence curves measured on and off the gold nanostructure, respectively, for the P780 sample with corresponding fits shown by the black-solid curves. The blue dotted curve showing the theoretically expected enhancement is obtained by fitting the blue data points (excluding the outlier) to Eq.~\eqref{eq:Lsattilde} with $\zeta$ as the only fitting parameter.}
\label{fig:RecordEnhancement}
\end{figure}

For all designs, we observe a drastic increase in the UCL enhancement by decreasing the excitation intensity (compare the upper and lower panels of Fig.~\ref{fig:Enh_Y}), indicating that the data are measured under the influence of saturation. To demonstrate this, we investigate the best-performing sample, P780, at the optimal conditions, i.e., \ang{1} angle of incidence and s-polarized excitation, at even lower excitation intensities. The laser power, and hence the intensity, is attenuated with neutral-density (ND) filters, thus maintaining the same laser-beam area as was the case for the results presented in the lower panels of Fig.~\ref{fig:Enh_Y}. In this case, we measured a maximum UCL enhancement factor of \num{913 \pm 51} (see blue open-faced circles in Fig.~\ref{fig:RecordEnhancement}) at an excitation intensity of \SI{1.7}{\watt\per\square\centi\meter}. The blue data point in Fig.~\ref{fig:RecordEnhancement} at high intensity and low UCL enhancements, measured at a similar angle of incidence and polarization but with the small laser-beam area taken from the upper panel of Fig.~\ref{fig:Enh_Y}b, is included to present all relevant data. The measurements indicate that even stronger enhancements could be measured by decreasing the excitation intensity even more, which is, however, impractical due to the signal-to-noise ratio of the integrating-sphere setup used.

Nevertheless, we have measured UCL enhancement factors between 30 and 913 for the same photonic structure at the same polarization and angle of incidence of the excitation source by merely varying the excitation intensity. This considerable span of enhancement factors demonstrates the arbitrariness in stating a UCL enhancement factor alone without explaining the proper experimental context. We wish to remove this arbitrariness, once and for all, such that it is possible to make a sensible quantification of the ability of a photonic structure to enhance the UCL. To achieve this, we shall define below a light-concentration factor $C_{\mathrm{ns}}$ as the proper quantification of the photonic structure. In order to get there, we remind the reader that the UCL enhancement is defined by the ratio $Y_{\mathrm{UCL}}^{\mathrm{on}}/Y_{\mathrm{UCL}}^{\mathrm{off}}$, where we know from Eq.~\eqref{eq:Ysat} that $Y_{\mathrm{UCL}}^{\mathrm{off}} \propto f(I/I_{\mathrm{sat}})$. It is thus tempting to investigate if we can find a similar relation for the UCL enhancement on the nanostructure. Therefore, we measure carefully the intensity dependence of the $Y_{\mathrm{UCL}}$ as exemplified in Fig.~\ref{fig:overview}c, both on and off the gold nanostructures at the same polarization and incidence angle as in the enhancement measurements just described. These intensity-dependence measurements are conducted without the use of an integrating sphere to improve the signal-to-noise ratio significantly, and the results are shown in Fig.~\ref{fig:RecordEnhancement} on (red circles) and off (red squares) the gold nanostructures. For both data sets, the solid black curves correspond to fits to Eq.~\eqref{eq:Ysat}. In other words, the experimental signals follow the model $Y_{\mathrm{UCL}}^{\mathrm{off}}(I) = A^{\mathrm{off}}f(I/I_{\mathrm{sat}}^{\mathrm{off}})$ and $Y_{\mathrm{UCL}}^{\mathrm{on}}(I) = A^{\mathrm{on}}f(I/I_{\mathrm{sat}}^{\mathrm{on}})$, where the fitted saturation intensities are $I_{\mathrm{sat}}^{\mathrm{on}} = \SI{0.59 \pm 0.05}{\watt\per\square\centi\meter}$ and $I_{\mathrm{sat}}^{\mathrm{off}} = \SI{39 \pm 11}{\watt\per\square\centi\meter}$, respectively\footnote{The difference in saturation intensities found off the nanostructure here and the value stated in Sec.~\ref{subsec:UCLMeas} are due to different polarization and incidence angle of the excitation laser}. The proportionality factors $A^{\mathrm{off}}$ and $A^{\mathrm{on}}$ depend on the experimental light collection efficiency, which is not the same on and off the nanostructures due to, among other, variations in the light-emission pattern.

While we expect from Ref.~\cite{Christiansen2019-1} that Eq.~\eqref{eq:Ysat} provides a good description for the intensity-dependence measurement off the nanostructures, it is quite surprising, at first glance, that a satisfactory fit can also be obtained on the nanostructures. One may be tempted to conclude that the approximation $\overline{f\left( x \right)} \approx f\left( \overline{x} \right)$ is valid even when the intensity distribution in the film is highly nonuniform under the influence of the nanostructures. However, in Sec.~4.2 of the supporting information, we argue that a better approximation is $\overline{ f\left(x \right)} \approx \zeta f \left(\overline{x}/\zeta \right)$, where $\zeta$ is a measure of the inhomogeneity of the intensity distribution inside the upconverting film and must be well-chosen in the range from 0.38 to 1. This fact explains why the UCL-intensity dependence on the gold nanostructure follow Eq.~\eqref{eq:Ysat}, and Sec.~4.2 of the supporting information also explains that the corresponding on-structure saturation intensity is given by
\begin{equation}
\label{eq:Isat_on}
I_{\mathrm{sat}}^{\mathrm{on}} = I_{\mathrm{sat}}^{\mathrm{off}}\zeta \cdot \frac{\overline{|\bm{E}_{\mathrm{b}}|^2}}{\overline{|\bm{E}|^2}} \equiv I_{\mathrm{sat}}^{\mathrm{off}}\zeta/C_{\mathrm{ns}},
\end{equation}
where the concentration factor $C_{\mathrm{ns}} \equiv \overline{|\bm{E}|^2}/\overline{|\bm{E}_\mathrm{b}|^2}$ of the mean EM energy density was defined. Hence, if we can determine the value of $\zeta$, it is possible in turn to calculate $C_{\mathrm{ns}}$ as a reliable, intensity-independent measure of the performance of the photonic structure. In order to achieve this, we show in Sec.~4.2 of the supporting information that the above-mentioned success of fitting the red data points in Fig.~\ref{fig:RecordEnhancement} to the $f$-function in Eq.~\eqref{eq:Ysat} immediately leads to the conclusion that the theoretically predicted enhancement in Eq.~\eqref{eq:Lsat} can be simplified to
\begin{equation}
\label{eq:Lsattilde}
\tilde{L}_{\mathrm{UCL}} = \zeta \frac{ f \left( \frac{I}{I_{\mathrm{sat}}^{\mathrm{on}}} \right) } { f \left( \frac{I}{I_{\mathrm{sat}}^{\mathrm{off}}} \right)},
\end{equation}
where $f \left( \frac{I}{I_{\mathrm{sat}}^{\mathrm{on}}} \right)$ and $f \left( \frac{I}{I_{\mathrm{sat}}^{\mathrm{off}}} \right)$ are exactly the functions known from the fitting in Fig.~\ref{fig:RecordEnhancement}. It should be noted that using the same polarization and incidence angle for all data points in Fig.~\ref{fig:RecordEnhancement} is required for this relation to be valid. Hence, to establish the value of $\zeta$, we only need to calibrate Eq.~\eqref{eq:Lsattilde} to the experimental UCL enhancement measurements. The dotted blue curve in Fig.~\ref{fig:RecordEnhancement} is obtained exactly in this way by fitting Eq.~\eqref{eq:Lsattilde} to the blue data points (excluding the outlier) while treating $\zeta$ as the only free parameter. With this, we have determined a reasonable $\zeta$-value of 0.48, well within the allowed range. The outlier is measured at such high excitation intensities that deviations from Eq.~\eqref{eq:Ysat} are expected to occur due to significant excitation to higher-energy states of \ce{Er^3+} not included in the model. This was previously observed for a core-shell \ce{NaYF_4}:\ce{Er} sample with a saturation intensity similar to the $I_{\mathrm{sat}}^{\mathrm{on}}$ stated above \cite{Christiansen2019-1}. With the $\zeta$-value at hand, we can now determine $C_{\mathrm{ns}} = \zeta I_{\mathrm{sat}}^{\mathrm{off}} /I_{\mathrm{sat}}^{\mathrm{on}} = \num{32 \pm 10}$. In comparison, the simulated concentration factor is moderately smaller at a value of \num{23}, consistent with the fact that the experimental UCL enhancements are somewhat higher than the simulation for P780 at these experimental conditions, see Fig.~\ref{fig:Enh_Y}b.

\section{Discussion}
\label{sec:Discussion}
Let us use our new understanding to consider the applicability of existing upconverters for solar-cell improvements under realistic excitation conditions, i.e., one sun. The introduced concentration factor, $C_{\mathrm{ns}}$, becomes important for two reasons: First, the total-absorption rate of an upconverting film is enhanced exactly by $C_{\mathrm{ns}}$ when any kind of photonic enhancement is present, see Eq.~(25) in Sec.~4.2 of the supporting information. Second, together with the $\zeta$-parameter the concentration factor can be interpreted as a lowering of the saturation intensity by the factor $\zeta/C_{\mathrm{ns}}$, and as a result, the saturation intensity approaching the ideal excitation condition of $I \approx 10I_{\mathrm{sat}}$, suggested in Ref.~\cite{Christiansen2019-1}. Nevertheless, the impressive $C_{\mathrm{ns}}$-factor reported in this article is still not enough for a working solar-cell device with currently available erbium-based upconverters. To arrive at this conclusion, consider that the available radiation energy from the sun in the absorption band of trivalent erbium is about \SI{3e-03}{\watt\per\centi\meter\squared}. This excitation intensity is roughly four orders of magnitude lower than the desired excitation regime for a \ce{TiO2}:\ce{Er} upconverter. Even if we had chosen the more efficient upconverter system of \ce{NaYF_4}:\ce{Er}, with a significantly lower saturation intensity \cite{Christiansen2019-1}, we are still between one and two orders of magnitude short in concentration factor. This estimate is even assuming that we could achieve similar photonic concentration in the \ce{NaYF4}, which is highly unlikely due to the smaller refractive index limiting the waveguiding efficiencies \cite{VesterPetersen2018}. Moreover, not much can be gained from a further photonic concentration since this must come at the expense of lowering the bandwidth of the photonic structure. A naive estimate on the bandwidth lowering is $\SI{1500}{\nano\meter}/C_{\mathrm{ns}} \approx \SI{50}{\nano\meter}$, which is already similar to the absorption bandwidth of \ce{Er^3+}. Instead, new materials could be introduced to improve performance, such as fluorescent concentrators, concentrating the entire available EM radiation from the sun in the range from \SIrange{1100}{1450}{\nano\meter} into the absorption band of \ce{Er^3+} as proposed in Ref.~\cite{Goldschmidt2008}. However, most desirable would be, also, to find new upconverting materials or hosts with larger absorption cross sections of the active material and ideally with high refractive index ensuring the capability of efficient waveguide coupling. After all, this will lower the saturation intensity as well as increase the total absorption rate, and thus significantly lower the demand for a high concentration of the EM radiation in large upconverting volumes.

\section{Conclusion}
\label{sec:conclusion}
In conclusion, we have designed and fabricated highly efficient photonic structures for enhancing the upconversion process in thin films of \ce{TiO2}:\ce{Er}. The UCL enhancement dependence on the unit-cell period of the nanostructure is studied through parametrically optimized designs in spired by the topology optimization. An unprecedented UCL enhancement factor of \num{913 \pm 51} has been measured at an excitation intensity of \SI{1.7}{\watt\per\square\centi\meter}. A model for the UCL enhancement is developed, which agrees reasonably with the measured UCL enhancements of the fabricated structures. The model further allows for an experimental determination of the concentration of the mean EM energy density in the upconverting film -- an essential parameter when assessing the applicability of photonic enhanced upconverters for solar-cell applications.

\section{Acknowledgments}
This work is supported by Innovation fund Denmark under the project "SunTune".

\bibliography{refMS}
\newpage



\section*{Supplementary material}

# Supporting information:
# Strongly enhanced upconversion in trivalent erbium ions by tailored gold nanostructures: toward high-efficient silicon-based photovoltaics

Jeppe Christiansen[a,1], Joakim Vester-Petersen[b,1], Søren Roesgaard[a,c], Søren H. Møller[a], Rasmus E. Christiansen[d], Ole Sigmund[d], Søren P. Madsen[b], Peter Balling[a,c], Brian Julsgaard[a,c]

[a]*Department of Physics and Astronomy, Aarhus University, Ny Munkegade 120, DK-8000 Aarhus C, Denmark*
[b]*Department of Engineering, Aarhus University, Inge Lehmanns Gade 10, DK-8000 Aarhus C, Denmark*
[c]*Interdisciplinary Nanoscience Center (iNANO), Aarhus University, Gustav Wieds Vej 14, DK-8000 Aarhus C, Denmark*
[d]*Department of Mechanical Engineering, DTU, Nils Koppels Allé 404, DK-2800 Kgs. Lyngby, Denmark*

**Contents**

---

[1]Contributted equally to this work

# 1. Numerical calculations

The period used in the topological optimization was chosen based on a simple wave-guide analysis to phase match a normal incident plane wave into a guided mode in the thin film of 320 nm-upconverting material [10]. As the gradient-based approach in topology optimization ensures convergence only to a local minimum, a multi-start procedure was used where a few optimizations were run with random initial material-distribution in the design domain, obtained by assigning each voxel a random $\rho$-value between 0 and 1. These calculations were optimizing the merit function (Eq. (4) of the main text) as a *min-max* optimization over s and p polarizations at the wavelength 1520 nm. A 10 nm-voxel size and first-order shape functions [5] in the finite element model were used for fast computation. An example of the iterations of such a calculation leading to a design similar to P780* is attached to the submission as a web-enhanced object, where $\rho \geq 0.1$ is shown in grayscale with black corresponding to $\rho = 1$. $\rho$ values less than 0.1 were not plotted. This design was then subsequently used in a high-resolution (5 nm voxels) *min-max* topology optimization considering both s and p polarizations at the wavelengths 1490 nm, 1500 nm, and 1510 nm to better match the structure to the experimental settings. This optimization led to the design P780* shown in Fig. 2 of the main text.

In the implementation of the topology optimization, the design variables essentially live in a 2D space (imagined to be on the top surface of the thin film), and they are then extruded into a 3D structure in the direction of the upward surface normal of the thin film, in order to limit the designs to only those compatible with the production. When realizing the design, the sides of the structure are expected to be slanted with an angle of 75°, as described in Sec. 3.2 of the main text. This slanting was introduced in the numerical design, and the unit cell was scaled to the measured periods when compared with the experiments in Fig. 3 of the main text. In addition, the accuracy of the numerical calculation was increased by using second-order shape functions in the finite element model, as opposed to the first-order shape functions used in the topological optimization.

The designs P780, P800, and P1000 were made by interpreting P780* as a central square and outer rings. The corners of the central square were rounded with a radius of 25 nm, corresponding roughly to the minimum radius achievable in the EBL production using the PEC method described in Ref. [4]. The sides of the structure were slanted with an angle of 75°. The merit function (Eq. (4) of the main text) was optimized using a *min-max* optimization for wavelengths 1490 nm, 1500 nm, and 1510 nm for the P780 design and 1480 nm, 1500 nm, and 1520 nm for the P800 and P1000 designs with the ring radius, ring thickness, and square side lengths as optimization variables. The minimum ring thickness was set to 45 nm, which is the minimum achievable thickness in the production. In all cases, the ring thickness was found to be 45 nm. Without the constraint on the ring thickness, a value between 20 and 40 nm were found in the different cases. As with P780*, the unit cells were also scaled to the measured periods when comparing to the experiments in, e.g., Fig. 3 of the main text.

The solution to the vector-wave equation (Eq. (3) of the main text) as well as the topology and parametric optimizations were conducted in the commercial software-package COMSOL MultiPhysics® [1]. For the topology optimization, the gradient-based optimization algorithm Globally Convergent Method of Moving Asymptotes (GCMMA) [9] implemented in COMSOL is used, and the entire topology-optimization procedure is controlled and automated using a custom MATLAB code interfacing with COMSOL through the LiveLink API. The parametric optimization of P780, P800, and P1000 is performed directly within COMSOL via the build-in derivative-free Nelder-Mead optimization algorithm.

# 2. Sample fabrication

## 2.1. Radio-frequency magnetron sputtering

Starting with a 0.5 mm-SiO$_2$ substrate, the TiO$_2$:Er was deposited using a radio-frequency magnetron sputtering system from AJA Orion ATC. The targets were commercially produced (by Able Targets) from powders of TiO$_2$ and Er$_2$O$_3$ at an erbium concentration of 5.1 at.%. The sputtering process was conducted in an argon atmosphere with 2 % oxygen at a pressure of 3 mmHg. The sputtering was done with fixed sputtering power of 100 W and with the substrate temperature fixed at 350 °C. These conditions were found

|   | P780* | P780 | P800 | P1000 |
|---|-------|------|------|-------|
| x | 782 (±2) | 781 (±3) | 798 (±4) | 1006 (±2) |
| y | 792 (±2) | 789 (±3) | 810 (±3) | 1015 (±1) |

Table 1: Average unit-cell period of the produced samples with the parentheses indicating the measurement uncertainty. All numbers are in nm.

to minimize unwanted nonradiative relaxation[6]. The deposition time was calibrated to achieve a film thickness of 320 nm. The fabricated thin films show no cyrstal structure.

### 2.2. Electron-beam lithography

The gold nanostructures were fabricated on the 320 nm-thick $TiO_2$:Er film in a $1.2 \times 1.2$ mm$^2$ area of the thin film surface to allow excitation on and off the the nano-structured area.

Prior to the EBL process, the samples were spin-coated with a positive resist (AR-P 6200.09) and post-baked at 150 °C for 60 s, yielding a 250 nm-thick film. Subsequently, a layer of 40 nm conductive polymer (AR-PC 5090) was spin coated and post-baked at 85 °C also for 60 s. The exposure was carried out using a FEI Magellan 400 SEM system at 30 kV accelerating voltage and 44 pA current. The SEM apparatus was equipped with a Raith pattern generator ($100 \times 100$ µm$^2$ writing field in 6 nm step size). Following exposure, the conductive polymer was removed by rinsing in deionized water for 60 s. Development of the positive resist was carried out for 60 s in a process resist (AR 600-546) and for 30 s in isopropanol as a stopper. Subsequently, layers of 3 nm Ti (adhesion layer) and 50 nm gold were deposited through the developed mask using a Polyteknik Cryofox Explorer 500 GLAD physical vapor-deposition system equipped with an e-beam-gun source (deposition rate 0.5 and 0.3 Å s$^{-1}$ for titanium and gold, respectively). Finally, the samples were soaked in remover (AR 300-76) overnight and sonicated for 180 s in order to remove unwanted material.

The optimal exposure pattern for the designs has been found using the "Proximity-effect correction" procedure described in Ref. [4].

## 3. Measurements

### 3.1. Optical-diffraction measurements

The average unit-cell dimensions of the fabricated structures were measured by optical diffraction using a helium-neon laser in a simple diffraction setup. The sample is illuminated from the quartz side by the helium-neon laser. A diffraction pattern is observed on a screen at a distance of around 1 m. From the distance to the screen, the distance from the 0'th diffraction order to the 1'st, and the wavelength of the helium-neon laser (632.8 nm), the period of the nanostructure is determined using Bragg's law. The results are stated in table 1.

### 3.2. Upconversion-luminescence measurements

The upconversion luminescence (UCL) was measured by exciting the samples with a 1500 nm-diode laser. The UCL intensity is recorded by an integrated Princeton Instrument Acton 2358 spectrograph and a PIXIS:100BR CCD camera. Example spectra with and without exciting the nanostructured area are shown in Fig. S1.

The polarization of the excitation laser is controlled by a Thorlabs half-wave plate (WPHSM05-1550) and a Thorlabs polarization filter (LPIREA100-C). The laser excitation intensity is controlled by utilizing two different beam areas, achieved by moving a focusing lens, and by attenuation with neutral-density filters. The UCL yield is determined by integrating the UCL intensity spectrum over the luminescence peak centered around 980 nm, see Fig. S1. The UCL enhancement, exemplified by the large difference in the two curves shown in Fig. S1, is computed as the ratio of the UCL yield when exciting on and off the gold nanostructures. The UCL-enhancement measurements were all carried out with the samples placed in a 150 mm-Labsphere integrating sphere to obtain identical collection efficiencies. The enhancement results with the horizontal

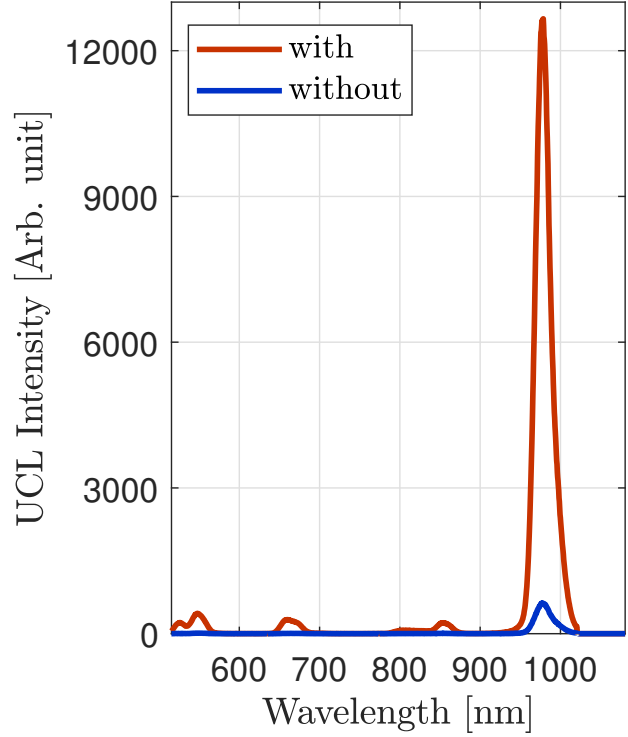

Figure S1: Spectra exemplifying both the upconversion luminescence with (red curve) and without (blue curve) gold nanostructures on the film surface.

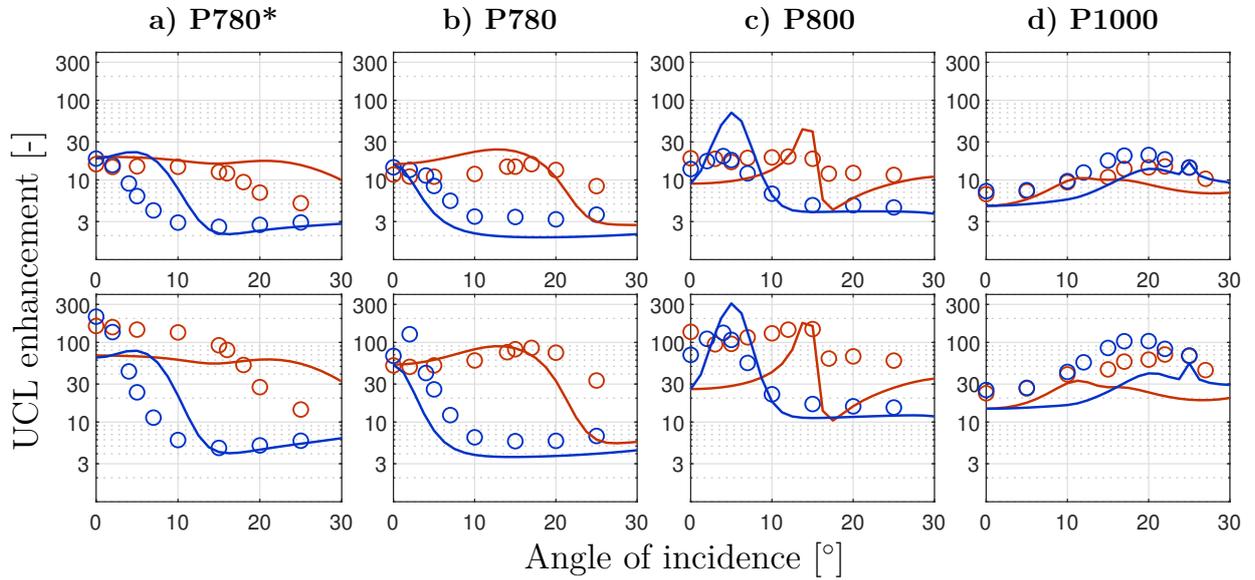

Figure S2: UCL enhancements plotted against the angle of incidence in colored circles of all four investigated samples with the horizontal period parallel with the electric field of the excitation source. In the upper row the "high" excitation intensity $323\,\mathrm{W\,cm^{-2}}$ is shown, with the "low" excitation intensity $5.8\,\mathrm{W\,cm^{-2}}$ underneath. The measured enhancement for p-polarized excitation is plotted in red, whereas the s-polarized measurements are plotted in blue. The solid line indicates the theoretical enhancement computed by the use of Eq. (2) of the main text.

4

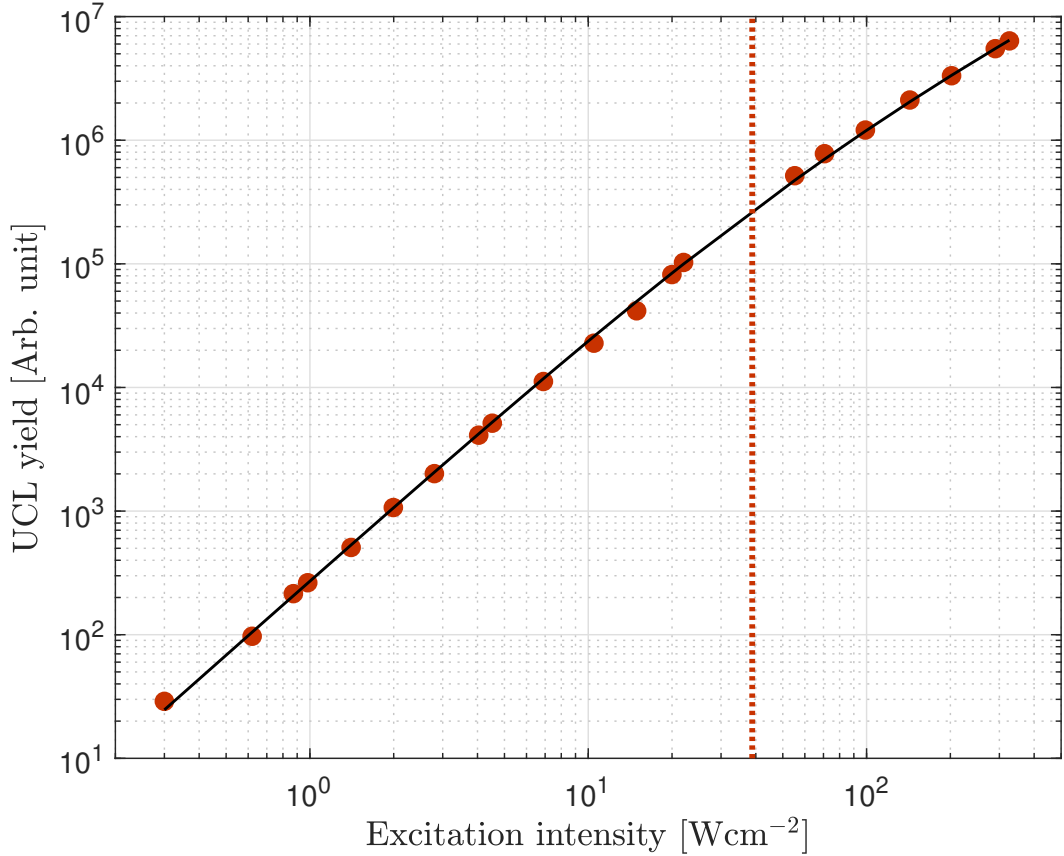

Figure S3: Intensity-dependence measurement for the bare $TiO_2$ film plotted with red dots with the corresponding fit to Eq. (1) cof the main text plotted as the solid black curve. The saturation intensities indicated with the dashed red vertical line is determined to be $20 \pm 6 \, \mathrm{W\,cm^{-2}}$, for the investigated case of p-polarized excitation under an 50° angle of incidence.

period (corresponding to x in Table 1) parallel to the electric field of the excitation laser for all samples are shown in Fig. S2, with the other case shown in Fig. 3 of the main text. The upconversion-luminescence intensity-dependence measurement is carried out similarly to the UCL enhancement measurements with the exceptions that the measurements were carried out without the use of an integrating sphere since the collection efficiency of the optical setup is not dependent on the excitation level. This allows for a better signal-to-noise ratio, which is required to investigate the UCL yield over a large dynamic range. The intensity-dependence is measured for a fixed angle of incidence of 50° and fixed p-polarized excitation. The saturation intensity, $I_{\mathrm{sat}}$ is determined by fitting the measurement to Eq. (1) of the main text. The results are shown in Fig. S3.

3.3. Beam-area estimation

Since a 1500 nm-diode laser is used as the excitation source, the beam cannot be imaged by a standard Si-CCD camera, which introduces some challenges when measuring the laser-beam area. We have measured the beam area by use of a movable razor blade and a power meter. The razor blade is placed where the beam area of the laser is to be measured, and the power meter somewhat after that. Assuming a Gaussian spatial intensity profile with the $x$ and the $y$ directions in the direction of the major and minor axes of the beam area, respectively, the intensity is given by: $I(x,y) = I_0 \exp\left[-2(x^2/w_x^2 + y^2/w_y^2)\right]$. By translating the

| | |
|---|---|
| Large beam area | $2.80 \pm 0.24 \times 10^{-3}$ cm$^2$ |
| Small beam area | $5.0 \pm 1.5 \times 10^{-5}$ cm$^2$ |

Table 2: Measured beam areas determined by the measurement and corresponding fits shown in Fig. S4.

razor blade along the $x$-direction, one can show that the measured power, $P_{\text{det}}$, is given by

$$P_{\text{det}}(x) = \frac{P_{\text{tot}}}{2} \left( 1 + \text{erf}\left( \frac{\sqrt{2}\,(x - x_0)}{w_x} \right) \right), \tag{1}$$

where $w_x$ and $w_y$ are the beam radii in the x and y-directions at the measured position in the beam path, and $P_{\text{tot}}$ is the total laser power. The beam area can thus be found by fitting Eq. (1) to data, see Fig. S4. From the maximum and minimum obtained beam waist the beam area, $\mathcal{A}$, are computed by $\mathcal{A} = \pi \min(w) \max(w)$ thus assuming an elliptical beam with $\min(w)$ and $\max(w)$ the major and minor axes, respectively.

To find the major a minor axes, the razor blade has been rotated in steps of 30° in the plane perpendicular to the propagation direction, for both the small and the large beam area used in the UCL measurements. The data and corresponding fits are shown in the upper panels of Fig. S4, with the corresponding beam radii in the lower panels. As seen in the upper panels, the data for the large beam area is nicely fitted whereas discrepancies are observed for the small beam area due to the spatial mode of the laser being only approximately Gaussian. Although unfortunate that we cannot provide a more accurate measure of the laser-beam area, it is important to note that the stated uncertainty reflects a systematic uncertainty for all intensities, whereas the relative uncertainty is much smaller. In other words, when comparing two different UCL measurements at the same intensity, there is rather large uncertainty in the exact value of the intensity used, reflected by the large, stated error, but the variation in intensity between the two measurements is much less. Therefore, the high uncertainty does not affect the uncertainty in the measured UCL enhancements, since these are independent of the excitation intensity as long as the UCL yields on and off the nanostructures are measured at similar conditions. The high uncertainty in the beam area, and thereby the intensities, will though affect the determined saturation intensities, since the absolute scale of the horizontal axes in Fig. S2 above and Fig. 3 of the main text are determined using the uncertain beam area.

### 3.4. Extinction cross section measurements

The extinction cross section has been measured in the range from 900 nm to 1800 nm using a LAMBDA-1050 UV/Vis spectrophotometer from Perkin Elmer Inc., see Fig. S5. The extinction cross section was calculated as the difference of the measured direct transmittance on and off the gold nanostructures, as explained in Ref. [8]. The simulated extinction cross section is computed as the difference in total transmittance in the presence and absence of gold nanostructures as computed by the FEM. Here, the total transmittance includes all light transmitted through the sample, whereas the direct transmittance only accounts for the part of the transmitted light that is not deflected. The comparison is thus not between exactly the same physical quantities, but the general trends will be similar. The good agreement between the simulated and measured extinction cross section spectra brings great confidence to the validity of the FEM calculations. Note also the narrow resonance at 1500 nm for P780* and P780, while it is slightly broader for the P800 and significantly broader for P1000. This agrees with the interpretation of a coupling dominated by waveguide effects for P780 and P780* and by plasmonic effects for P1000 as described in Sec. 4 of the main text.

## 4. The upconversion model

### 4.1. Derivation of the saturation model for upconversion enhancement

The response of the upconverting Er$^{3+}$ ions to electromagnetic (EM) radiation follows a set on nonlinear differential equations, as explained in Ref. [2]. The rate of upconversion emission, $\Gamma_{\text{UCL}}$, is a measure of the

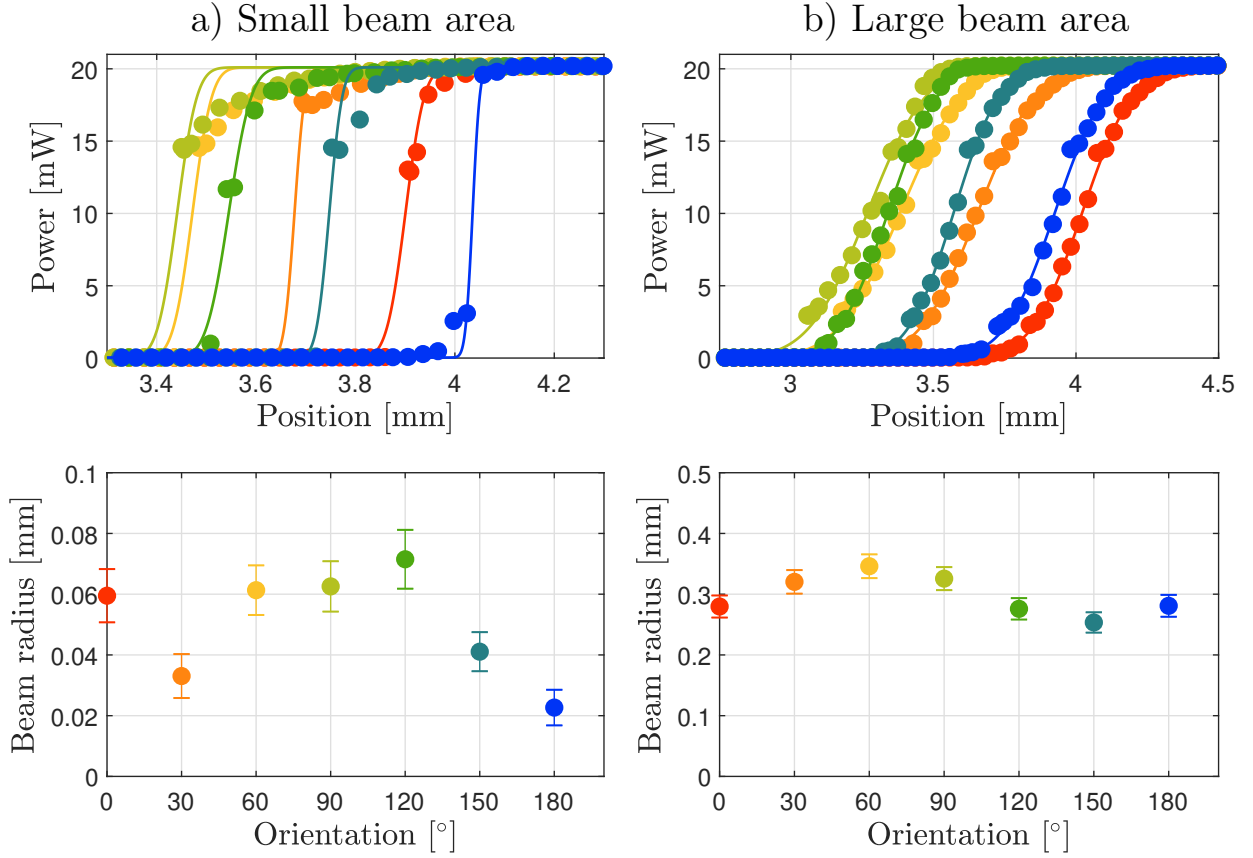

Figure S4: In the left panels (a), data for a hard focus of the laser-beam spot corresponding to a small beam area, and the corresponding data for the weak focus and large beam area to the right (b). In the upper panels, the plotted points are measured power of the laser as a function of the position of the razor blade, and the solid curves are fits to Eq. (1). In the lower panels, the fitted beam radii are plotted against the orientation of the scan in steps of 30°. The resulting beam area, $\mathcal{A}$, is computed as the product $\mathcal{A} = \pi \min(w_i) \max(w_i)$.

number of emitted photons per second per $\text{Er}^{3+}$ ion, and following Eq. (9) in Ref. [2], this can be expressed as:

$$\Gamma_{\text{UCL}} = \Gamma_{\text{eff}} \cdot f\left(\frac{|\mathbf{E}(x,y,z)|^2}{|E_{\text{sat}}|^2}\right), \tag{2}$$

where $\Gamma_{\text{eff}}$ is a constant to be discussed below, $\mathbf{E}$ is the complex amplitude of the electric field at the location of the $\text{Er}^{3+}$, $E_{\text{sat}}$ is a material-dependent parameter describing the characteristic field at saturation to be discussed later, and $f$ is a function given by

$$f(\mu) = 1 - \sqrt{1+\mu} + \frac{\mu}{2}. \tag{3}$$

The constant $\Gamma_{\text{eff}}$ is given by

$$\Gamma_{\text{eff}} = \frac{A_{31}}{\Gamma_{31}} \frac{\sigma_{12}\epsilon_0 c n_{\text{R}} |E_{\text{sat}}|^2}{2h\nu}, \tag{4}$$

where $A_{31}$ and $\Gamma_{31}$ are the spontaneous emission rate and total decay rate, respectively, on the upconversion transition $3 \to 1$ (corresponding to the $^4I_{11/2} \to {}^4I_{15/2}$ transition at around 980 nm in $\text{Er}^{3+}$), $\sigma_{12}$ is the absorption cross section on the transition $1 \to 2$ (corresponding to the $^4I_{15/2} \to {}^4I_{13/2}$ transition at around 1500 nm in $\text{Er}^{3+}$), $\epsilon_0$ is the vacuum permittivity, $c$ is the speed of light in vacuum, $n_{\text{R}}$ is the real part of the

7

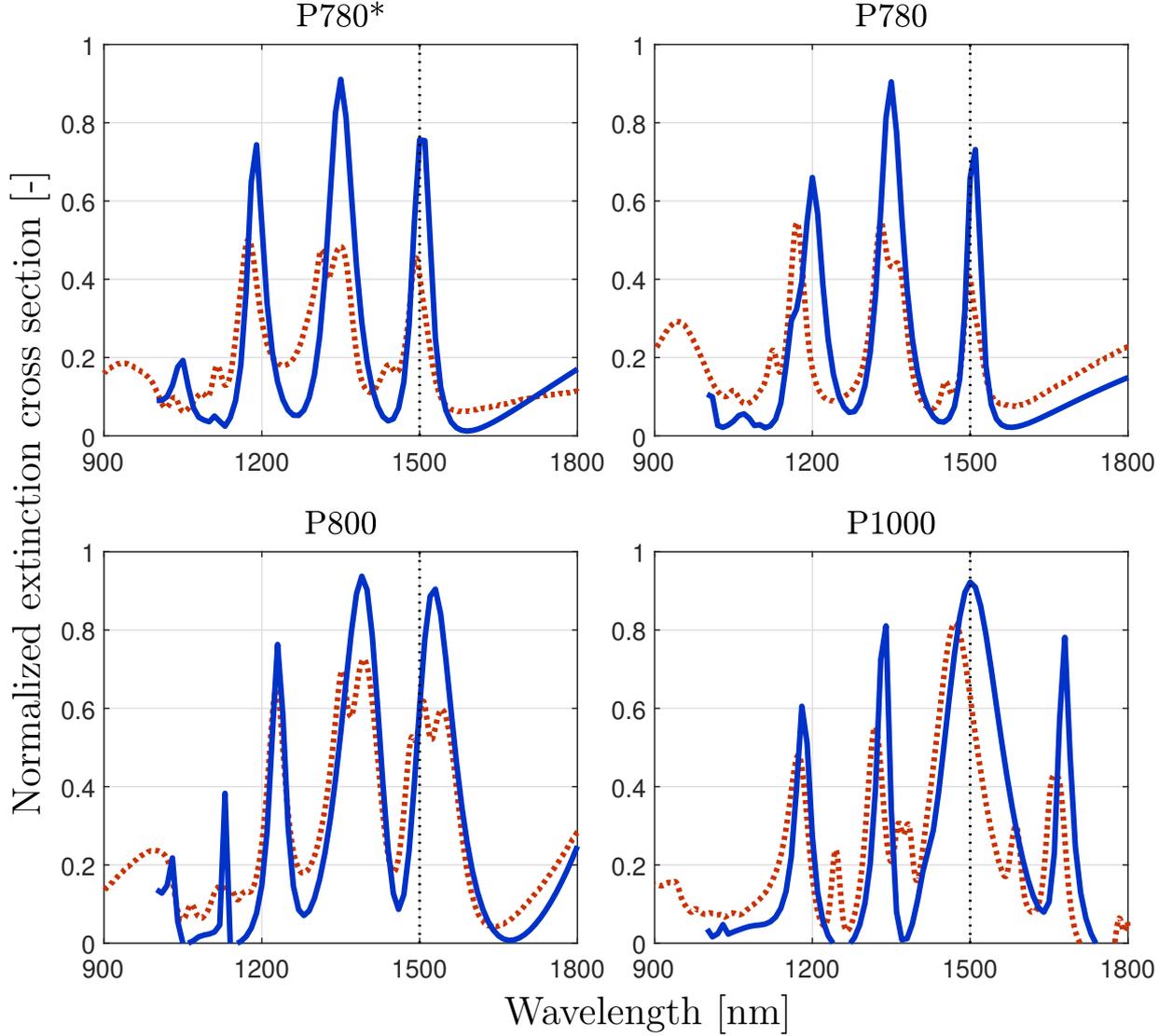

Figure S5: Measured (red dashed curve) and simulated (blue solid curve) extinction cross section of all investigated samples.

refractive index, $h$ is Planck's constant, and $\nu$ is the frequency of the (monochromatic) EM radiation. It should be noted that the mathematical formalism in Ref. [2] is expressed in terms of the incoming intensities, which is here translated into electric field amplitudes $E$ according to: Intensity $= \frac{1}{2}\epsilon_0 n_\mathrm{R} c |E|^2$ with thin-film interference accounted for by the FEM simulations. $|E_\mathrm{sat}|^2$ is hereby related to the material parameters via

$$|E_\mathrm{sat}|^2 = \frac{2}{\epsilon_0 n_\mathrm{R} c} \frac{h\nu \Gamma_{21}(W_\mathrm{cr} + \Gamma_{43})}{8\sigma_{12} W_\mathrm{etu} \Gamma_{43}}, \qquad (5)$$

with $\Gamma_{21}$ and $\Gamma_{43}$ the total decay rates on the upconversion transition $2 \to 1$ and $4 \to 3$, respectively (corresponding to the $^4I_{13/2} \to {}^4I_{15/2}$ and $^4I_{9/2} \to {}^4I_{11/2}$ transitions in $\mathrm{Er}^{3+}$), $W_\mathrm{etu}$ is the energy-transfer process $(2,2) \to (4,1)$ between two excited $\mathrm{Er}^{3+}$, and $W_\mathrm{cr}$ is the inverse process to $W_\mathrm{etu}$ often denoted cross relaxation.

The upconversion yield, $Y_\mathrm{UCL}$, describes the total number of emitted upconverted photons per second

8

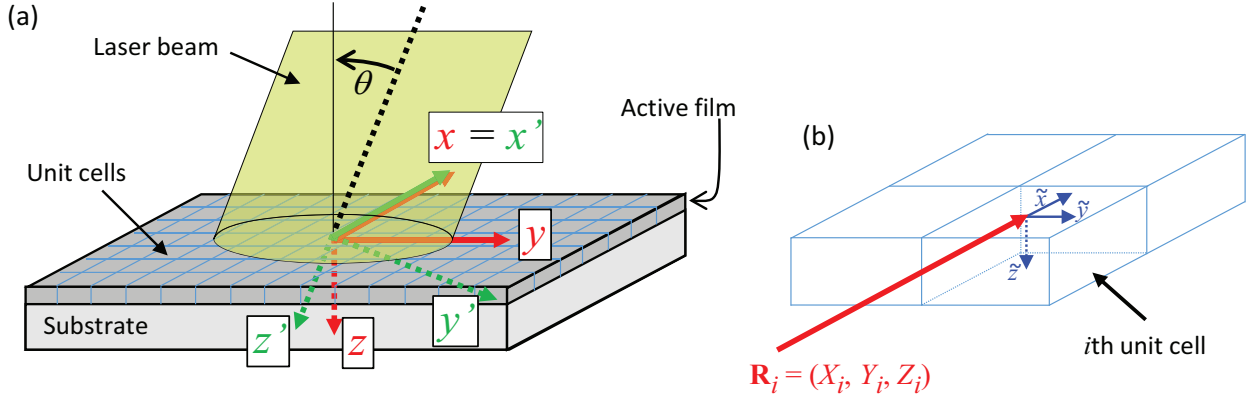

Figure S6: (a) Coordinate systems on the macroscopic scale: The system $(x, y, z)$, shown in red, follows the directions of the active film, while a rotated system $(x', y', z')$, shown in green, follows the incoming laser beam, which is incident under an angle $\theta$. The origo of both systems is defined as the center of the laser spot on the film surface. The active film may be patterned by nanostructures in a periodic manner with a spatial repetition period indicated by the unit cells. (b) Coordinate systems on the local scale: The vector $\mathbf{R}_i$, shown in red, points to the $i$th unit cell, whereas the system $(\tilde{x}, \tilde{y}, \tilde{z})$, shown in blue, parametrizes the volume of a single unit cell.

and is found by integrating $\Gamma_{\text{UCL}}$ over the entire active film (see Fig. S6(a)):

$$Y_{\text{UCL}} = N \int_{\text{film}} \Gamma_{\text{UCL}} \, dx \, dy \, dz \approx N\Gamma_{\text{eff}} \int_{\text{film}} f\left(\frac{|\mathbf{E}(x,y,z)|^2}{|E_{\text{sat}}|^2}\right) dx \, dy \, dz, \qquad (6)$$

where $N$ is the density of $Er^{3+}$ ions. In the second step, it is assumed that $\Gamma_{\text{eff}}$ has no spatial dependence. In principle, the quantum efficiency of the light emission, and hence $\Gamma_{\text{eff}}$, can be affected by the presence of metallic nanostructures. However, this effect is only significant in the close vicinity of the nanostructures where the UCL is typically quenched [3], and the approximation is very reasonable for the 320 nm-thick film investigated here.

The electric field $\mathbf{E}(x, y, z)$ has a spatial dependence for several reasons: (i) the active film may be patterned by nanostructures, (ii) there may be thin-film interference in the active film, and (iii) the incoming laser beam has a spatial dependence of its complex amplitude given by:

$$|E_{\text{laser}}(x', y', z')|^2 = \frac{4P}{c\epsilon_0 \pi w^2} \exp\left(-\frac{2(x'^2 + y'^2)}{w^2}\right). \qquad (7)$$

This expression follows from some algebraic manipulation and the expressions for Gaussian laser beams in the paraxial approximation [7]. The coordinates $(x', y', z')$ refer to a coordinate system aligned with the laser beam [see Fig. S6(a)], $P$ is the total power of the laser beam, and $w$ is the beam radius which is assumed to be constant. In principle, $w$ will vary along the line of propagation, but for a moderately focused laser beam, these variations take place on a much longer scale than the thickness of the investigated film. Along the same lines of thought, the spatial extent of the laser spot is assumed to be much larger than the spatial period (unit cells) of the nanostructures on the active film. This will be used in the following to obtain approximate and tractable ways of calculating the upconversion yield, $Y_{\text{UCL}}$.

The two coordinate systems, $(x, y, z)$ and $(x', y', z')$ shown in Fig. S6(a), are related to each other via the angle of incidence $\theta$:

$$\begin{aligned} x' &= x, \\ y' &= y \cos\theta + z \sin\theta, \\ z' &= -y \sin\theta + z \cos\theta. \end{aligned} \qquad (8)$$

9

In addition, let the vector $\mathbf{R}_i = (X_i, Y_i, Z_i)$ point to the $i$th unit cell of the active film and define a local coordinate system $(\tilde{x}, \tilde{y}, \tilde{z})$ within one unit cell [see Fig. S6(b)]. The modulus square of the electric field amplitude in the $i$th unit cell can then conveniently be written as:

$$|\mathbf{E}_i(x,y,z)|^2 = |\mathbf{E}(X_i + \tilde{x}, Y_i + \tilde{y}, Z_i + \tilde{z})|^2. \tag{9}$$

In the presence of nanostructures, this electric field varies rapidly with $(\tilde{x}, \tilde{y}, \tilde{z})$, while at the same time, the incident laser field has variations over a much larger scale and is rather constant over a few unit cells. This motivates an approximation: For the $i$th unit cell, assume that the incoming laser field is a plane wave and use periodic boundary conditions for a numerical calculation of $\mathbf{E}_i(x, y, z)$ since the neighboring unit cells experience essentially the same incoming field. This simplifies the calculations significantly, and it is only necessary to carry out a numerical determination of $\mathbf{E}_i(x, y, z)$ for a single unit cell. In this calculation, let $E_0$ be the complex amplitude of the incoming plane wave, incident at the angle $\theta$ and with a polarization described by $p$. The resulting electric field within the unit cell in the active film is then denoted as $\mathbf{E}(\tilde{x}, \tilde{y}, \tilde{z}, \theta, p)$. Since $E_0$ can be chosen arbitrarily in the calculation, we shall by $\tilde{\mathbf{E}}$ denote the simulated normalized electric field relative to the incoming amplitude:

$$\tilde{\mathbf{E}}(\tilde{x}, \tilde{y}, \tilde{z}, \theta, p) = \frac{\mathbf{E}(\tilde{x}, \tilde{y}, \tilde{z}, \theta, p)}{E_0}. \tag{10}$$

In experiment, the magnitude of the incoming plane wave is set by the Gaussian laser beam for the $i$th unit cell:

$$|E_0|^2 = |E_{\text{laser}}(X_i', Y_i', 0)|^2, \tag{11}$$

where it is sufficient to evaluate Eq. (7) only at the center of the unit cell. Hence, the electric field within the $i$th unit cell is given by:

$$|\mathbf{E}_i(x,y,z)|^2 = |\tilde{\mathbf{E}}(\tilde{x}, \tilde{y}, \tilde{z}, \theta, p)|^2 \frac{4P}{c\epsilon_0 \pi w^2} \exp\left(-\frac{2(X_i'^2 + Y_i'^2)}{w^2}\right). \tag{12}$$

In other words, the single-unit-cell numerical result for the relative electric field within the film, denoted as $|\tilde{\mathbf{E}}(\tilde{x}, \tilde{y}, \tilde{z}, \theta, p)|^2$ and parametrized by the local-scale coordinates $(\tilde{x}, \tilde{y}, \tilde{z})$, is simply multiplied by the actual field amplitude of the incoming Gaussian laser beam at the location of the $i$th unit cell, which is parametrized by the macro-scale coordinates $(X_i', Y_i', Z_i')$. The integral in Eq. (6) then becomes simpler; a sum over unit cells on the macro scale and an integration over local coordinates within one unit cell:

$$\begin{aligned} Y_{\text{UCL}} &= N\Gamma_{\text{eff}} \sum_i \int_{\text{cell}} f\left(\frac{|\tilde{\mathbf{E}}(\tilde{x},\tilde{y},\tilde{z},\theta,p)|^2}{|E_{\text{sat}}|^2} \frac{4P}{c\epsilon_0 \pi w^2} \exp\left(-\frac{2(X_i'^2 + Y_i'^2)}{w^2}\right)\right) \,d\tilde{x}\,d\tilde{y}\,d\tilde{z} \\ &= \frac{N\Gamma_{\text{eff}}}{\Delta X' \Delta Y'} \int_{\text{cell}} \left\{\sum_i f\left(\frac{|\tilde{\mathbf{E}}|^2}{|E_{\text{sat}}|^2} \frac{4P}{c\epsilon_0 \pi w^2} \exp\left(-\frac{2(X_i'^2 + Y_i'^2)}{w^2}\right)\right) \Delta X' \Delta Y'\right\} d\tilde{x}\,d\tilde{y}\,d\tilde{z} \\ &\approx \frac{N\Gamma_{\text{eff}}}{\Delta X' \Delta Y'} \int_{\text{cell}} \left\{\int f\left(\frac{|\tilde{\mathbf{E}}|^2}{|E_{\text{sat}}|^2} \frac{4P}{c\epsilon_0 \pi w^2} \exp\left(-\frac{2(X'^2 + Y'^2)}{w^2}\right)\right) dX'\,dY'\right\} d\tilde{x}\,d\tilde{y}\,d\tilde{z} \\ &= \frac{N\Gamma_{\text{eff}}}{\Delta X' \Delta Y'} \int_{\text{cell}} \left\{\int_0^\infty f\left(2\beta \exp\left(-\frac{2R'^2}{w^2}\right)\right) 2\pi R' \,dR'\right\} d\tilde{x}\,d\tilde{y}\,d\tilde{z}. \end{aligned} \tag{13}$$

The second step left out $(\tilde{x}, \tilde{y}, \tilde{z})$ for brevity and introduced $\Delta X' \Delta Y'$ in the numerator and denominator, where $\Delta X'$ is the distance between $X_i'$ coordinates for two neighbouring unit cells along the $x'$-axis and likewise for $\Delta Y'$. Note that from Eq. (8) it follows that $\Delta X' = \Delta X$ and $\Delta Y' = \Delta Y \cos\theta$, where the unprimed $\Delta X$ and $\Delta Y$ refer to the widths of the unit cell along the $x$ and $y$ axes. The third step above exploits the fact that $X_i'$ and $Y_i'$ vary slowly on the scale of $\Delta X'$ and $\Delta Y'$, recognizing that the sum in the curly parentheses can be approximated as an integral. The fourth step is a change to polar coordinates for the integral in the curly brackets, $R'^2 = X'^2 + Y'^2$, and an abbreviation is introduced within the $f$-function:

$$\beta = \frac{|\tilde{\mathbf{E}}(\tilde{x},\tilde{y},\tilde{z},\theta,p)|^2}{|E_{\text{sat}}|^2} \frac{2P}{c\epsilon_0 \pi w^2}. \tag{14}$$

10

Now, define a new parameter

$$\xi = 2\beta \exp\left(-\frac{2R'^2}{w^2}\right), \qquad (15)$$

which is an injective function of $R'$ and can be used to integrate the expression in the curly brackets in Eq. (13) by substitution. With $d\xi = -4R'\xi dR'/w^2$ and new integration limits $2\beta$ and $0$, one finds:

$$\begin{aligned}\int_0^\infty f\left(\beta \exp\left(-\frac{2R'^2}{w^2}\right)\right) 2\pi R'\, dR' &= \frac{\pi w^2}{2}\int_0^{2\beta} \frac{f(\xi)}{\xi} d\xi \\ &= \frac{\pi w^2}{2}\int_0^{2\beta} \left(\frac{1-\sqrt{1+\xi}}{\xi} + \frac{1}{2}\right) d\xi = \pi w^2 f_\text{gauss}(\beta),\end{aligned} \qquad (16)$$

where the function $f_\text{gauss}$ is given by

$$f_\text{gauss}(\beta) = 1 + \ln\left(\frac{\sqrt{1+2\beta}+1}{2}\right) + \frac{\beta}{2} - \sqrt{1+2\beta}. \qquad (17)$$

The $f$-function from Eq. (3) and the above $f_\text{gauss}$-function are identical to the curly brackets of Eqs. (10) and (14) in Ref. [2] and shown in Fig. 2(a) of that reference. Inserting Eq. (16) in Eq. (13) leads to the upconversion yield

$$\begin{aligned}Y_\text{UCL} &= \frac{Nd\pi w^2 \Gamma_\text{eff}}{\Delta X \Delta Y d \cos\theta}\int_\text{cell} f_\text{gauss}(\beta)\, d\tilde{x}\, d\tilde{y}\, d\tilde{z} \\ &= \frac{Nd\pi w^2 \Gamma_\text{eff}}{\Delta X \Delta Y d \cos\theta}\int_\text{cell} f_\text{gauss}\left(\frac{|\tilde{\mathbf{E}}(\tilde{x},\tilde{y},\tilde{z},\theta,p)|^2}{|E_\text{sat}|^2}\frac{2P}{c\epsilon_0 \pi w^2}\right) d\tilde{x}\, d\tilde{y}\, d\tilde{z} \\ &= \frac{Nd\pi w^2 \Gamma_\text{eff}}{\cos\theta}\overline{f_\text{gauss}\left(\frac{|\tilde{\mathbf{E}}(\tilde{x},\tilde{y},\tilde{z},\theta,p)|^2}{|E_\text{sat}|^2}\frac{2P}{c\epsilon_0 \pi w^2}\right)}.\end{aligned} \qquad (18)$$

In the first line, the thickness, $d$, of the active film was introduced, and in the second line, $\beta$ was simply inserted according to Eq. (14). Remembering that $\Delta X \Delta Y d$ is the volume of the thin-film under the unit cell, the bar above $f_\text{gauss}$ in the third line denotes a spatial average over the unit cell. We also note that $I = P/\pi w^2$ is the characteristic intensity of the incoming Gaussian laser beam.

It remains to determine the value of $E_\text{sat}$, i.e., the characteristic field amplitude within the active film at which the $Er^{3+}$ ions reach their saturation point. Ref. [2] approached the same problem in an operational way in terms of the incoming intensity by defining $I_\text{sat}$; the value of the incoming intensity at which the $Er^{3+}$ ions reach their saturation point. In that reference, the upconversion yield, $Y_\text{UCL}$, was measured as a function of incoming intensity $I$, and $I_\text{sat}$ was derived as a fitting parameter. Such measurements are typically carried out for a specific angle of incidence, $\theta_s$, and a chosen polarization, $p_s$, with the subscript "s" referring to "saturation measurement". To see how the parameter $I_\text{sat}$ derived by this method relates to the $E_\text{sat}$, one should first acknowledge that for a given incoming characteristic intensity, $I$, the electric field $\mathbf{E}(x,y,z)$ inside the active film is slightly varying due to thin-film interference. A sensible measure of $E_\text{sat}$ is found by requiring that the average of $|\mathbf{E}|^2$ inside the film is equal to $|E_\text{sat}|^2$ when $I$ is equal to what we denote as $I_\text{sat}$. Since an intensity $I$ and a complex field amplitude $E_0$ are related by $I = \frac{1}{2}\epsilon_0 c |E_0|^2$ in free space, the factor $\frac{2I_\text{sat}}{c\epsilon_0}$ must describe the incoming $|E_0|^2$ at the saturation point. In this case, the field inside the unit cell must be given by:

$$|\mathbf{E}_s(x,y,z)|^2 = |\tilde{\mathbf{E}}(\tilde{x},\tilde{y},\tilde{z},\theta_s,p_s)|^2 \frac{2I_\text{sat}}{c\epsilon_0}, \qquad (19)$$

where $\tilde{\mathbf{E}}_s = \tilde{\mathbf{E}}(\tilde{x},\tilde{y},\tilde{z},\theta_s,p_s)$ is the numerically determined relative field inside the active film for the angle and polarization corresponding to the experimental settings of the saturation measurement. We thus find:

$$|E_\text{sat}|^2 = \overline{|\tilde{\mathbf{E}}_s|^2}\frac{2I_\text{sat}}{c\epsilon_0}, \qquad (20)$$

11

which can be inserted into Eq. (18), and one finds the final expression for the upconversion yield:

$$Y_{\text{UCL}} = \frac{Nd\pi w^2 \Gamma_{\text{eff}}}{\cos\theta} \overline{f_{\text{gauss}}\left(\frac{|\tilde{\mathbf{E}}(\tilde{x},\tilde{y},\tilde{z},\theta,p)|^2}{|\tilde{\mathbf{E}}_{\text{s}}|^2}\frac{I}{I_{\text{sat}}}\right)}. \tag{21}$$

We stress that $\tilde{\mathbf{E}}$ and $\tilde{\mathbf{E}}_{\text{s}}$ both represent numerically calculated electric fields. The field, $\tilde{\mathbf{E}}(\tilde{x},\tilde{y},\tilde{z},\theta,p)$, in the numerator in the equation above can in principle show very large spatial variations if metal nanoparticles are present on the active film. For this reason the averaging, denoted by the overline, must be performed outside the $f_{\text{gauss}}$-function. However, if we choose a typical experimental setting for determination of the saturation intensity, i.e., in the absence of nanostructures, we must set $\tilde{\mathbf{E}} = \tilde{\mathbf{E}}_{\text{s}}$. Since this field has small spatial variations in practice, the upconversion yield during the saturation measurement reduces to

$$\begin{aligned}Y_{\text{UCL,s}} &= \frac{Nd\pi w^2 \Gamma_{\text{eff}}}{\cos\theta} \overline{f_{\text{gauss}}\left(\frac{|\tilde{\mathbf{E}}_{\text{s}}|^2}{|\tilde{\mathbf{E}}_{\text{s}}|^2}\frac{I}{I_{\text{sat}}}\right)} \\ &\approx \frac{Nd\pi w^2 \Gamma_{\text{eff}}}{\cos\theta} f_{\text{gauss}}\left(\frac{\overline{|\tilde{\mathbf{E}}_{\text{s}}|^2}}{|\tilde{\mathbf{E}}_{\text{s}}|^2}\frac{I}{I_{\text{sat}}}\right) = \frac{Nd\pi w^2 \Gamma_{\text{eff}}}{\cos\theta} f_{\text{gauss}}\left(\frac{I}{I_{\text{sat}}}\right),\end{aligned} \tag{22}$$

which, apart from the $\cos\theta$-scaling of the beam area $\pi w^2$ due to the arbitrary angle of incidence $\theta$, corresponds exactly to Eq. (14) in Ref. [2] that was used to fit the experimental saturation curves of Fig. 4 in that reference.

With this derivation, it is finally possible to derive an expression for the upconversion luminescence enhancement, i.e., the ratio between the upconversion yield in the presence and the absence of metal nanostructures on the active film. If the numerically simulated relative field is denoted by $\tilde{\mathbf{E}}$ in the presence of nanostructures and by $\tilde{\mathbf{E}}_{\text{b}}$ in the absence of nanostructures (subscript "b" for "background"), the upconversion enhancement must be given by:

$$\text{Enhancement} = \overline{f_{\text{gauss}}\left(\frac{|\tilde{\mathbf{E}}|^2}{|\tilde{\mathbf{E}}_{\text{s}}|^2}\frac{I}{I_{\text{sat}}}\right)} \Big/ \overline{f_{\text{gauss}}\left(\frac{|\tilde{\mathbf{E}}_{\text{b}}|^2}{|\tilde{\mathbf{E}}_{\text{s}}|^2}\frac{I}{I_{\text{sat}}}\right)} \tag{23}$$

This is equivalent to Eq. (2) of the main text (for brevity, the subscript "gauss" is removed there), and this is used to calculate the model curves of Fig. 3 in the main text. It should be stressed that these calculations rely only on the experimental material parameter $I_{\text{sat}}$ and on dielectric functions of the various materials (gold nanostructures, $TiO_2$:Er upconverter, fused quartz substrate, and air) constituting the physical system.

### 4.2. Absorption, light concentration, and quantum efficiency

Turning to the absorption process, we define $Y_{\text{abs}}$ as the total number of absorbed photons per second. This can be calculated as an integral over the photon-absorption rate per volume, $NR_{12}$, where the absorption rate per $Er^{3+}$ ion is

$$R_{12} = \frac{\sigma_{12}\epsilon_0 c n_{\text{R}} |\mathbf{E}|^2}{2h\nu}. \tag{24}$$

We thus find:

$$\begin{aligned}Y_{\text{abs}} &= \int_{\text{film}} R_{21} N \, \mathrm{d}\tilde{x}\, \mathrm{d}\tilde{y}\, \mathrm{d}\tilde{z} \\ &= \frac{N\sigma_{12}\epsilon_0 c n_{\text{R}}}{2h\nu} \sum_i \int_{\text{cell}} |\tilde{\mathbf{E}}(\tilde{x},\tilde{y},\tilde{x},\theta,p)|^2 \frac{4P}{c\epsilon_0 \pi w^2} \exp\left(-\frac{2(X_i'^2 + Y_i'^2)}{w^2}\right) \mathrm{d}\tilde{x}\, \mathrm{d}\tilde{y}\, \mathrm{d}\tilde{z} \\ &= \frac{2N\sigma_{12} n_{\text{R}} P}{h\nu \Delta X' \Delta Y' \pi w^2} \int_{\text{cell}} |\tilde{\mathbf{E}}(\tilde{x},\tilde{y},\tilde{x},\theta,p)|^2 \left\{\int \exp\left(-\frac{2(X'^2 + Y'^2)}{w^2}\right) \mathrm{d}X'\mathrm{d}Y'\right\} \mathrm{d}\tilde{x}\, \mathrm{d}\tilde{y}\, \mathrm{d}\tilde{z} \\ &= \frac{Nd\sigma_{12} n_{\text{R}} P}{h\nu \cos\theta} \frac{1}{\Delta X \Delta Y d} \int_{\text{cell}} |\tilde{\mathbf{E}}(\tilde{x},\tilde{y},\tilde{x},\theta,p)|^2 \mathrm{d}\tilde{x}\, \mathrm{d}\tilde{y}\, \mathrm{d}\tilde{z} \\ &= \frac{Nd\sigma_{12} n_{\text{R}} P}{h\nu \cos\theta} \cdot \overline{|\tilde{\mathbf{E}}(\tilde{x},\tilde{y},\tilde{x},\theta,p)|^2},\end{aligned} \tag{25}$$

where the mathematical step is similar to the procedure above for the calculation of $Y_{\text{UCL}}$. It should be noted that the absorption rate is proportional to the square of the electric field inside the film, and if we define the concentration factor $C_{\text{ns}} = \overline{|\tilde{\mathbf{E}}|^2}/|\tilde{\mathbf{E}}_b|^2$ as the relative increase in this squared electric field due to the presence of the nanostructures, the total absorption rate is simply enhanced by this concentration factor.

With the upconversion yield and total absorption rate at hand, it is now possible to calculate the internal upconversion quantum yield as the ratio of the number of emitted upconverted photons per second to the number of absorbed photons per second:

$$\text{IUCQY} = \frac{Y_{\text{UCL}}}{Y_{\text{abs}}} = \frac{h\nu\pi w^2 \Gamma_{\text{eff}}}{\sigma_{12} n_R P \overline{|\tilde{\mathbf{E}}|^2}} \cdot \overline{f_{\text{gauss}}\left(\frac{|\tilde{\mathbf{E}}|^2}{|\tilde{\mathbf{E}}_s|^2}\frac{\bar{I}}{I_{\text{sat}}}\right)} \tag{26}$$

$$= \frac{A_{31}}{\Gamma_{31}}\left(\frac{\overline{|\tilde{\mathbf{E}}_s|^2}}{|\tilde{\mathbf{E}}|^2}\frac{I_{\text{sat}}}{\bar{I}}\right) \overline{f_{\text{gauss}}\left(\frac{|\tilde{\mathbf{E}}|^2}{|\tilde{\mathbf{E}}_s|^2}\frac{\bar{I}}{I_{\text{sat}}}\right)},$$

where Eqs. (4) and (20) was used in the second step. We note that this result has a striking similarity to Eq. (16) in Ref. [2], which is simply adjusted by the relative strength between the calculated electric fields in presence of nanostructures and under the circumstances of the calibration of the saturation intensity. We remind that the averaging over the unit cell in the final factor must be performed outside the function $f_{\text{gauss}}$ due to the nonlinear nature of the response of the $\text{Er}^{3+}$ to the radiation field. Nonetheless, in practice it is possible to fit an experimentally obtained upconversion luminescence yield *on the nanostructures* to a fitting function on the form $f_{\text{fit}} = A f_{\text{gauss}}(\bar{I}/I_{\text{sat}}^{\text{on}})$, where $I_{\text{sat}}^{\text{on}}$ and $A$ are fitting parameters, see Fig. 4 of the main text. For this reason, it is worthwhile to consider the possibility of obtaining an approximate expression for $\overline{f_{\text{gauss}}}$. Later, we shall argue that $\overline{f_{\text{gauss}}(\beta)} \approx \zeta f_{\text{gauss}}(\bar{\beta}/\zeta)$ is a reasonable approximation for a well-chosen value of $\zeta$, where $\beta = \frac{|\tilde{\mathbf{E}}|^2}{|\tilde{\mathbf{E}}_s|^2}\frac{\bar{I}}{I_{\text{sat}}}$ is the statistically distributed argument of the $f_{\text{gauss}}$ function. With this approximation at hand, the IUCQY can be simplified to:

$$\text{IUCQY} = \frac{A_{31}}{\Gamma_{31}}\frac{I_{\text{sat}}^{\text{on}}}{\bar{I}} f_{\text{gauss}}\left(\frac{\bar{I}}{I_{\text{sat}}^{\text{on}}}\right), \tag{27}$$

where

$$I_{\text{sat}}^{\text{on}} = I_{\text{sat}} \cdot \frac{\overline{|\tilde{\mathbf{E}}_s|^2}}{|\tilde{\mathbf{E}}_b|^2} \cdot \frac{\zeta}{C_{\text{ns}}} \tag{28}$$

is the characteristic saturation intensity, which is lowered mainly by the factor $C_{\text{ns}}$ due to the enhanced absorption and further slightly lowered by $\zeta < 1$ (see below) due to the nonlinear response of the upconversion process and the distribution of $|\tilde{\mathbf{E}}|^2$. The final factor $\overline{|\tilde{\mathbf{E}}_s|^2}/|\tilde{\mathbf{E}}_b|^2$ simply accounts for the fact that background electric field $\tilde{\mathbf{E}}_b$ "without" the nanostructures present and the field $\tilde{\mathbf{E}}_s$ during the saturation calibration may be different due to differences in incidence angle and polarization. Note that Eq. (27) is identical to Eq. (16) in Ref. [2] and thus constituting an effectively unified description.

In the absence of nanostructures, the electric field reduces to the background field, $\tilde{\mathbf{E}} \to \tilde{\mathbf{E}}_b$ such that $C_{\text{ns}} \to 1$ and $\zeta \approx 1$ (since the background field is nearly uniform, see below). We then define the "off" saturation intensity as $I_{\text{sat}}^{\text{off}} = I_{\text{sat}}\overline{|\tilde{\mathbf{E}}_s|^2}/|\tilde{\mathbf{E}}_b|^2 = I_{\text{sat}}^{\text{on}} C_{\text{ns}}/\zeta$, and the exact enhancement calculation from Eq. (23) can be approximated as

$$\text{Enhancement} \approx \zeta \frac{f_{\text{gauss}}\left(\frac{\bar{I}}{I_{\text{sat}}^{\text{on}}}\right)}{f_{\text{gauss}}\left(\frac{\bar{I}}{I_{\text{sat}}^{\text{off}}}\right)} \tag{29}$$

which has the asymptotic values of $C_{\text{ns}}$ for $\bar{I} \gg I_{\text{sat}}^{\text{off}}$ and $C_{\text{ns}}^2/\zeta$ for $\bar{I} \ll I_{\text{sat}}^{\text{on}}$. Let us now justify the approximation $\overline{f_{\text{gauss}}(\beta)} \approx \zeta f_{\text{gauss}}(\bar{\beta}/\zeta)$, which led to the above simplifications. To this end, we consider the sample P780 and calculate the exact value $\overline{f_{\text{gauss}}(\beta)}$ as a function of the incoming intensity $\bar{I}$ through the

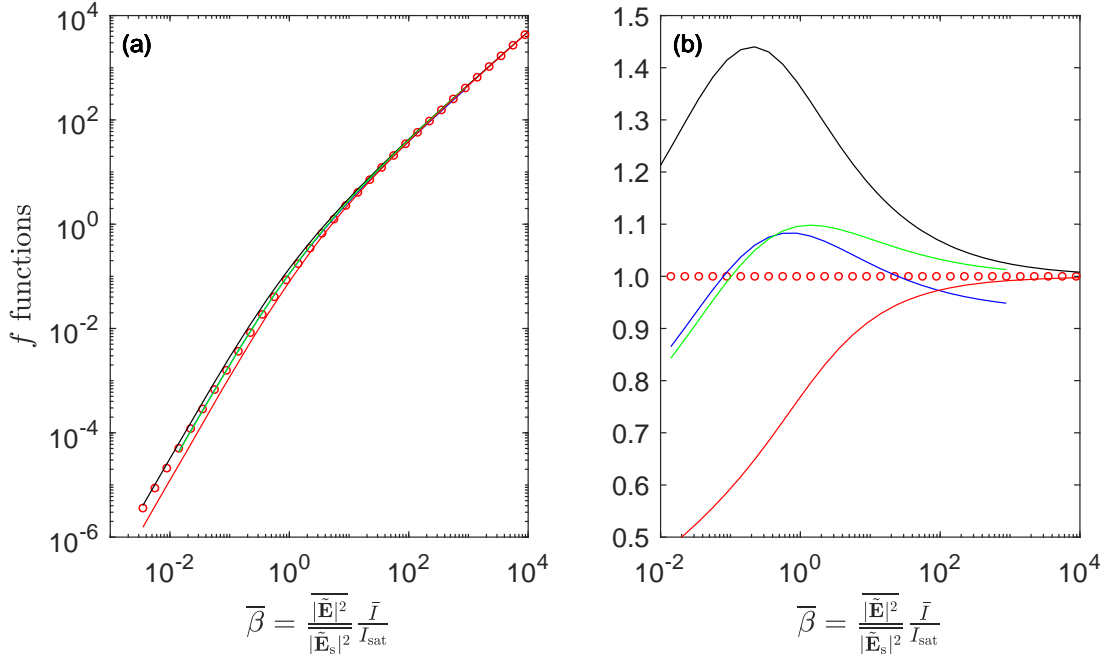

Figure S7: The two panels show the same curves; on a real scale in panel (a) and normalized to the exact function $\overline{f_{\text{gauss}}(\beta)}$ in panel (b). The different curves are: $\overline{f_{\text{gauss}}(\beta)}$ (red circles); $f_{\text{gauss}}(\overline{\beta})$ (red curve); $\zeta f_{\text{gauss}}(\overline{\beta}/\zeta)$ with asymptotically matched $\zeta = \overline{(|\tilde{\mathbf{E}}|^2)}^2/\overline{|\tilde{\mathbf{E}}|^4} = 0.38$ (black curve); $\zeta f_{\text{gauss}}(\overline{\beta}/\zeta)$ with a freely fitted $\zeta = 0.53$ (green curve); a general UCL fit $A f_{\text{gauss}}(\overline{\beta}/\zeta)$ with $A = 0.48$ and $\zeta = 0.52$ (blue curve).

parameter $\overline{\beta}$ (defined above). The result is shown by the red circles in Fig. S7, an approximate expressions should be as close to this result as possible. In comparison, the red solid curve shows the neglect of the field distribution corresponding to $\overline{f_{\text{gauss}}(\beta)} \rightarrow f_{\text{gauss}}(\overline{\beta})$. At high intensities, this substitution is exact since $f_{\text{gauss}}(\beta)$ depends linearly on $\beta$ in that limit, and taking the average inside or outside the $f_{\text{gauss}}$ function does thus not matter. In contrast, for very small intensities $f_{\text{gauss}}$ scales quadratically with $\beta$, and for this reason $f_{\text{gauss}}(\overline{\beta})/\overline{f_{\text{gauss}}(\beta)} = \overline{(|\tilde{\mathbf{E}}|^2)}^2/\overline{|\tilde{\mathbf{E}}|^4} \equiv \zeta_{\text{asymp}}$ when $\beta \ll 1$. Since the variance of $|\tilde{\mathbf{E}}|^2$ can be expressed as $\sigma^2(|\tilde{\mathbf{E}}|^2) = \overline{|\tilde{\mathbf{E}}|^4} - \overline{(|\tilde{\mathbf{E}}|^2)}^2$, the relative standard deviation of the energy density within the film must be: $\frac{\sigma(|\tilde{\mathbf{E}}|^2)}{|\tilde{\mathbf{E}}|^2} = \sqrt{\frac{1}{\zeta_{\text{asymp}}} - 1} \geq 0$, meaning that $\zeta_{\text{asymp}} \leq 1$. Hence, $f_{\text{gauss}}(\overline{\beta})$ must underestimate the exact calculation $\overline{f_{\text{gauss}}(\beta)}$, which is clearly seen by the red curve in Fig. S7(b). The construction $\overline{f_{\text{gauss}}(\beta)} \approx \zeta f_{\text{gauss}}(\overline{\beta}/\zeta)$ preserves the correct asymptotic limit when $\beta \gg 1$ and imposes the condition $\zeta f_{\text{gauss}}(\overline{\beta}/\zeta) \rightarrow \frac{\zeta_{\text{asymp}}}{\zeta} f(\beta)$ when $\beta \ll 1$. Hence, by setting $\zeta = \zeta_{\text{asymp}}$, the two asymptotic limits will be correctly reproduced as indicated by the black curves in Fig. S7. Evidently, we have in this case sacrificed an accurate reproduction of values around the most interesting regime near the saturation point. For this reason, the best approximation is obtained if $\zeta$ attains a compromising value such that $\zeta_{\text{asymp}} < \zeta < 1$. The green curve in Fig. S7 represents such a case, where $\zeta$ is determined as a free fitting parameter and the relative mistake is only of the order of 10 % in the broad vicinity of the saturation point. The blue curve represents a slightly more free fit $A f_{\text{gauss}}(\overline{\beta}/\zeta)$ with essentially the same $\zeta$ parameter as was obtained by the green curve. In Fig. 4 of the main text $\zeta$ is fitted to UCL enhancement data for the P780 sample using the approximate enhancement expression of Eq. (29) with $I_{\text{sat}}^{\text{on}}$ and $I_{\text{sat}}^{\text{off}}$ fixed at the fitted values. A reasonable $\zeta$-value of 0.48 within the allowed range was found yielding a concentration factor of $C_{\text{ns}} = 32 \pm 10$.

14



\end{document}